\documentclass[a4paper, 10 pt, conference]{ieeeconf}
\IEEEoverridecommandlockouts                              
\renewcommand{\baselinestretch}{0.95}

\usepackage{comment}
\usepackage{lettrine}
\usepackage{amsmath,amssymb}
\usepackage[ruled,vlined]{algorithm2e}
\usepackage{tikz}
\usepackage{pgfplots}
\usepackage{mathtools}
\usepackage{comment}
\usepackage{theorem}
\usepackage{standalone}
\usepackage{bbm}
\usepackage{graphicx}
\usepackage{caption}
\usepackage{subcaption}
\usepackage{dsfont}
\usepackage[font=footnotesize,labelfont=bf]{caption}
\usepackage{multirow}
\usepackage{lettrine}
\usepackage[noadjust]{cite}
\usepackage{units}

\usetikzlibrary{arrows,positioning,shapes,backgrounds,fit}

\newcommand{\dsqbL}{[\kern-0.15em[}
\newcommand{\dsqbR}{]\kern-0.15em]}

\pgfplotsset{compat=1.14}
\usetikzlibrary{arrows}
\newcommand{\be}[1]{\begin{equation}\label{#1}}
\newcommand{\benon}{\begin{equation*}}  
\newcommand{\bemuln}[1]{\begin{multline}\label{#1}}
\newcommand{\bemul}{\begin{multline*}}
\newcommand{\bea}{\begin{eqnarray*}}
\newcommand{\eea}{\end{eqnarray*}}
\newcommand{\been}[1]{\begin{eqnarray}\label{#1}}
\newcommand{\eeen}{\end{eqnarray}}
\newcommand{\began}[1]{\begin{gather}\label{#1}}
\newcommand{\bega}{\begin{gather*}}
\newcommand{\bealn}[1]{\begin{align}\label{#1}}
\newcommand{\beal}{\begin{align*}}
\newcommand{\bealatn}[2]{\begin{alignat}{#1}\label{#2}}
\newcommand{\bealat}{\begin{alignat*}}
\newcommand{\bexalatn}[1]{\begin{xalignat}\label{#1}}
\newcommand{\bexalat}{\begin{xalignat*}}

\theoremstyle{plain}

\def\bc{{\mathbf c}}

\def\bg{{\mathbf g}}

\def\bt{{\mathbf t}}

\def\bw{{\mathbf w}}
\def\bx{{\mathbf x}}

\def\texitem#1{\par\smallskip\noindent\hangindent 25pt
               \hbox to 25pt {\hss #1 ~}\ignorespaces}

\newcommand{\scrA}{\mathcal{A}}

\newcommand{\scrG}{\mathcal{G}}

\newcommand{\scrV}{\mathcal{V}}
\newcommand{\scrW}{\mathcal{W}}

\graphicspath{ {./fig/} }

\title{\LARGE \bf Congestion-aware Routing and Rebalancing of Autonomous Mobility-on-Demand Systems in Mixed Traffic*}

\author{Salom\'{o}n Wollenstein-Betech$^{1}$, Arian Houshmand$^1$,  Mauro Salazar$^2$, \\ Marco Pavone$^2$,  Christos G. Cassandras$^1$, and Ioannis Ch. Paschalidis$^1$
\vspace{-0.1cm}
\thanks{*This work was supported in part by NSF under grants ECCS-1509084, DMS-1664644, CNS-1645681, IIS-1914792, and CMMI-1454737, by AFOSR under grant FA9550-19-1-0158, by ARPA-E's NEXTCAR program under grant DEAR0000796, by the MathWorks, by the ONR under grant N00014-19-1-2571, by the NIH under grant 1R01GM135930, and by the Toyota Research Institute (TRI).
This article solely reflects the opinions and conclusions of its authors and not NSF, TRI, or any other entity. We thank D. Sverdlin-Lisker, Dr. I. New and Dr. K. Solovey for proofreading this paper.}
\thanks{$^{1}$ The authors are with the
Division of Systems Engineering, Boston University, Brookline, MA 02446 USA {\tt\small \{salomonw, arianhm, cgc, yannisp\}@bu.edu}}
\thanks{{$^{2}$ The authors are with the department of Aeronautics and Astronautics, Stanford University, Stanford, CA 94325 USA {\tt\small \{samauro,pavone\}@stanford.edu}}}
}

\begin{document}    
    \maketitle
    \thispagestyle{empty}
    \pagestyle{empty}
    
    \begin{abstract}\label{sec: abstract}
This paper studies congestion-aware route-planning policies for Autonomous Mobility-on-Demand (AMoD) systems, whereby a fleet of autonomous vehicles provides on-demand mobility under mixed traffic conditions.
Specifically, we first devise a network flow model to optimize the AMoD routing and rebalancing strategies in a congestion-aware fashion by accounting for the endogenous impact of AMoD flows on travel time.
Second, we capture reactive exogenous traffic consisting of private vehicles selfishly adapting to the AMoD flows in a user-centric fashion by leveraging an iterative approach.
Finally, we showcase the effectiveness of our framework with two case-studies considering the transportation sub-networks in Eastern Massachusetts and New York City. Our results suggest that for high levels of demand, pure AMoD travel can be detrimental due to the additional traffic stemming from its rebalancing flows, while the combination of AMoD with walking or micromobility options can significantly improve the overall system performance.
\end{abstract}

    \section{Introduction} \label{sec:intro}
\lettrine{I}{n the} past decade, the rapid adoption of smartphone technologies and wireless communications coupled with the emergence of sharing economies has resulted in a widespread use of Mobility-on-Demand (MoD) services.
One of the main operational challenges that these services face is deciding routing and rebalancing policies for their vehicles. Currently, MoD systems use \emph{user-centric} routing services (e.g., Waze and Google Maps) to route their vehicles, and dynamic pricing combined with a real-time heat-map of the users' demand to rebalance their fleets.

Given this \emph{user-centric} approach to route vehicles, in which every driver acts selfishly to minimize their own travel time, the network reaches an equilibrium known as the \emph{Wardrop} equilibrium~\cite{wardrop1952road}.  Unfortunately, these equilibria are in general suboptimal compared to the system optimum, achievable when the vehicles are coordinated by a central controller in a \emph{system-centric} fashion.
\begin{figure}[t]
	\centering
	\includegraphics[width=1\linewidth]{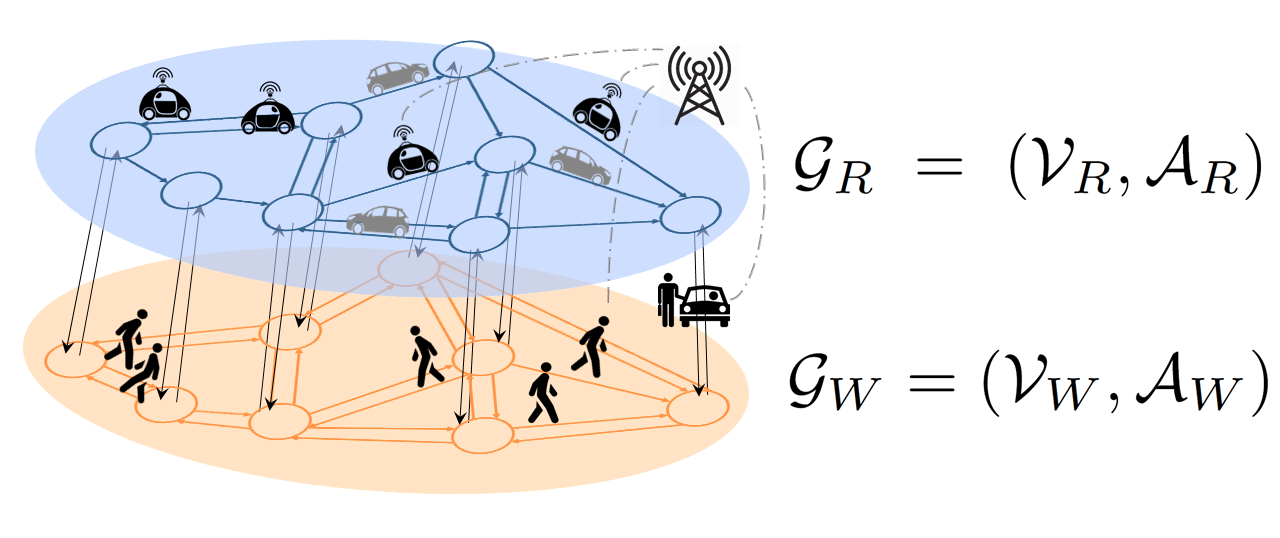}
	\caption{AMoD network (supergraph) consisting of two digraphs for the road (blue) and the walking (orange) network; the black arrows represent switching arcs. AMoD vehicles are in black and private vehicles in grey.}
	\label{fig:road-walking-graph}
	\vspace{-.45cm}
\end{figure}

Recently, the combination of MoD services with Connected and Automated Vehicles (CAVs) has attracted the interest of academia and industry, giving rise to Autonomous Mobility-on-Demand (AMoD) systems (see Fig.~\ref{fig:road-walking-graph}). These fleets of CAVs providing on-demand mobility are expected to reduce labor costs, accidents, harmful emissions~\cite{guanetti2018control}, and increase the efficiency of the fleets' operation as they can be \emph{centrally} controlled. Considering high penetration rates of AMoD in the mobility ecosystem, the routing and rebalancing policies designed to centrally control the vehicles will affect the congestion levels and, in turn, the routing decisions of privately owned vehicles.
In this context, this paper studies system-optimal routing and rebalancing strategies for AMoD systems in mixed-traffic conditions.

\emph{Related literature:}
AMoD systems and rebalancing policies have been extensively studied using simulation models~\cite{swaszek2019load,HoerlRuchEtAl2018,LevinKockelmanEtAl2017}, queuing-theoretical models~\cite{ZhangPavone2016,IglesiasRossiEtAl2016}, and network-flow models~\cite{PavoneSmithEtAl2012,RossiZhangEtAl2017}. In~\cite{swaszek2019load}, the rebalancing of an AMoD system is addressed using a data-driven real-time parametric controller. Alternatively, in~\cite{PavoneSmithEtAl2012}, the rebalancing problem is studied using a steady-state fluid model.
Although~\cite{swaszek2019load} and~\cite{PavoneSmithEtAl2012} seek to find effective rebalancing policies, they do not consider the impact of the AMoD routes on congestion, but rather assume travel times on the road links to be constant.

Little work has been done to solve the congestion-aware routing and rebalancing problem jointly. Most approaches leverage approximations of the travel time function relating traffic density to travel times to address the non-convex nature of the problem.
The authors of~\cite{RossiZhangEtAl2017} use a threshold model to show that under relatively mild assumptions rebalancing vehicles do not lead to an increase in congestion, suggesting that the joint problem can be decoupled without having a substantial negative impact on the solution's quality. Moreover, \cite{SalazarTsaoEtAl2019} introduced a piecewise-affine approximation of the travel time function in order to relax the problem to a quadratic program.
Yet, depending on the congestion levels, both approaches may lack in accuracy. Moreover, \cite{RossiZhangEtAl2017} and \cite{SalazarTsaoEtAl2019} assume a static exogenous traffic flows that does not change for varying AMoD routes. 
Finally, reactive private traffic was modeled in~\cite{houshmand2019penetration} to show that under a system-centric optimal-routing strategy both CAVs and non-CAVs can achieve better performance in terms of travel time and energy savings. However, such an approach neither captures rebalancing effects nor intermodal routing possibilities. 

\emph{Statement of contribution:} 
This paper bridges the gap between~\cite{SalazarTsaoEtAl2019} and~\cite{houshmand2019penetration}. Specifically, we study how system-optimal routing of AMoD services can affect the system-level performance in \emph{mixed-traffic} (presence of  AMoD and private vehicles in the road network). 
Similar to \cite{SalazarTsaoEtAl2019}, we assume that AMoD users can use multiple modes of transportation, i.e., autonomous taxi rides and walking. In addition, we assume the private vehicle flow to be \emph{reactive}, meaning that private vehicles will choose their routes selfishly considering the congestion stemming from the AMoD flow. To this end, we use the framework previously developed in~\cite{houshmand2019penetration} for modeling the interaction between AMoD and private vehicles. Moreover, we devise an approximation of the travel time function that is more accurate than the one proposed in~\cite{SalazarTsaoEtAl2019}, whilst still maintaining the quadratic convex structure of the AMoD problem.
The proposed model can efficiently compute congestion-aware routing and rebalancing strategies for a given demand and road network topology.
Finally, with this framework at hand, we analyze the trade-offs between the benefits of system-centric routing and the cost of rebalancing, and
investigate the achievable benefits stemming from the combination of AMoD with walking and micromobility options.

\emph{Organization:} The rest of the paper is organized as follows: In Section~\ref{sec:problem formulation} we provide  preliminaries of the model and its formulation. In Section~\ref{sec:travel-latency-function-approximation} we develop a convex approximation of the original problem to overcome its non-convex nature. We present experiments using the Eastern Massachusetts and New York City road networks in Section~\ref{sec:experiments}. Finally, in Section~\ref{sec:conclusions} we conclude the paper and point to future research directions.

\emph{Notation:} All vectors are column vectors and denoted by bold lowercase letters. We use ``prime'' to denote transpose, and use $\mathds{1_x}$ to denote the indicator function.

    \section{Problem Formulation} \label{sec:problem formulation}
In this section, we present mesoscopic models for planning the routing and rebalancing strategies used throughout the paper. First, we introduce the notation and preliminaries of transportation modeling. With this in hand, we model the system-centric routing and rebalancing of AMoD, followed by the user-centric model for private vehicles. Finally, we formulate the joint problem of congestion-aware routing and rebalancing of AMoD in mixed traffic.

\subsection{Preliminaries} \label{subsec:multi-class formu}
Consider an AMoD system which provides mobility services through two modes of transportation: walking and autonomous taxi-rides. To model the system, let $\scrG$ be a network (supergraph) composed of two layers, a road and a walking network. We denote by $\scrG_\mathrm{R}=(\scrV_\mathrm{R}, \scrA_\mathrm{R})$ the road network and by $\scrG_\mathrm{W}=(\scrV_\mathrm{W}, \scrA_\mathrm{W})$ the pedestrian graph where $(\scrV_\mathrm{R}, \scrA_\mathrm{R})$ and $(\scrV_\mathrm{W}, \scrA_\mathrm{W})$  are the sets of intersections (vertices) and streets (arcs) in the road and in the pedestrian network, respectively.
Then, the supergraph $\scrG=(\scrV, \scrA)$ is composed of $\scrG_\mathrm{R}$ and $\scrG_\mathrm{W}$, and a set of \emph{switching} arcs $\scrA_\mathrm{S} \subset \scrV_\mathrm{R} \times \scrV_\mathrm{W} \cup \scrV_\mathrm{W} \times \scrV_\mathrm{R}$ that connect the pedestrian and the road network layers to allow AMoD users to change modes (see Fig. \ref{fig:road-walking-graph}). Formally $\scrG$ is composed of the set of vertices $\scrV = \scrV_\mathrm{R} \cup \scrV_\mathrm{W}$ and arcs $\scrA = \scrA_\mathrm{R} \cup \scrA_\mathrm{W} \cup \scrA_\mathrm{S}$.

In order to model the demanded trips, let $\bw = (w_s, w_t)$ denote an Origin-Destination (OD) pair and $d_{\bw} \ge 0$ the demand rate at which customers request service per unit time from origin $w_s$ to destination $w_t$. Let $W$ be the total number of OD pairs and $\scrW = \{ \bw_k: \bw_k = (w_{sk}, w_{tk}), k = \{1,...,W\} \}$ the set of OD pairs. Let a vectorized version of the demand be $\bg=(d^{\bw}: \bw\in \scrW)$, which denotes the demand flows for all OD pairs.

To keep track of AMoD users' flow on an arc, we let $x_{ij}^{\bw}$ denote the AMoD flow induced by OD pair $\bw$ in link $(i,j)\in\scrA$. Given that the AMoD needs to rebalance its vehicles to ensure service, we let $x^r_{ij}$ be the \emph{rebalancing flow} on road $(i,j)$. Finally, to consider the interaction between the AMoD provider and the other vehicles, we let $x^p_{ij}$ be the  self-interested \emph{private vehicle} flow on $(i,j)$. We use the term \emph{private} as we assume that self-interested users must arrive at their destination with their vehicle and do not have the option of switching transportation mode (i.e., walking). 
To simplify notation, we let the AMoD user flow on any edge (road, walking, or switching) to be
\begin{equation} \label{eq:customer-carrying-flow}
    x^u_{ij} = \sum_{\bw \in \scrW} x^\bw_{ij}, \quad \quad \forall (i,j) \in \scrA,
\end{equation}  
and the total flow on a link to be
\begin{equation} \label{eq:total-flow}
    x_{ij}=x^u_{ij}+x^r_{ij}+x^p_{ij}, \quad \quad \forall (i,j) \in \scrA.
\end{equation}
Note that neither rebalancing flow $\bx^r$, nor private vehicle flow $\bx^p$ should exist on the switching arcs $\scrA_\mathrm{S}$ or walking arcs $\scrA_\mathrm{W}$. Hence, for those arcs we set $x^r_{ij}=x^p_{ij}=0, \ \forall (i,j) \in \scrA_\mathrm{S} \cup \scrA_\mathrm{W}$.

Let $t_{ij}(x) : \mathbb{R}_{+}^{| \scrA|} \mapsto \mathbb{R}_{+}$ be the \emph{travel time} function, i.e., the time it takes to cross link $(i,j)$ given the flow on that link. Using the same function structure as in \cite{Beckmann1955}, we characterize $t_{ij}$ as a function of the flow $x_{ij}$ with
\begin{equation}
    t_{ij}(x_{ij}) = t_{ij}^0f(x_{ij}/m_{ij}), \label{eq:latency-function}
\end{equation}
where $m_{ij}$ is the road's capacity, $f(\cdot)$ is a strictly increasing, positive, and continuously differentiable function, and $t_{ij}^0$ is the free-flow travel time on link $(i,j)$. We would like to consider functions with $f(0)=1$, which ensures that if there is no flow on the link, the travel time $t_{ij}$ is equal to the free-flow travel time.
Typically, travel time functions used by urban planners and researchers are polynomials which are hard to estimate \cite{wollenstein2019joint}. A widely used function is the {\em Bureau of Public Roads (BPR)} travel time function \cite{BPR1964} denoted by
\begin{equation} \label{bpr}
t_{ij}(x_{ij}) = t_{ij}^0(1+0.15(x_{ij}/m_{ij})^4).
\end{equation}
Throughout this paper, we use this function to decide the routes of AMoD users and private vehicles, given the network flow levels. For AMoD users who walk, we consider a constant travel time (independent of the flow) on each link.

\subsection{System-centric Routing and Rebalancing of AMoD} \label{subsec:routing-and-rebalancing}
Recall that our goal is to find the system-centric congestion-aware routes and rebalancing policy of an AMoD provider. The objective consists of minimizing the cost composed of the overall travel time of AMoD users, and a regularizer penalizing rebalancing flow. 

We formulate the problem similar to~\cite{SalazarTsaoEtAl2019} where we address it from an AMoD perspective.  Let $d^u_\bw$ be customer requests to the AMoD provider traveling from origin $w_s$ to destination $w_t$.  The problem we aim to solve is then expressed by
\begin{subequations} \label{eq:problem-formulation}
    \begin{flalign} 
        & \hspace{-.3cm} \min\limits_{\{\bx^{\bw}\}_{\bw \in \scrW} ,\bx^r}  J(\bx) :=  \sum\limits_{(i,j)\in \scrA} t_{ij}( x_{ij} ) x^u_{ij}  + \bc' \bx^r \label{eq:joint-obj} \\[3pt]
        & \hspace{.5cm}\text{s.t.} \sum\limits_{i:(i,j) \in \scrA} \hspace{-0.2cm} x^\bw_{ij} + \mathds{1}_{j=w_s} d^u_\bw = \hspace{-0.4cm} \sum\limits_{k:(j,k) \in \scrA} \hspace{-0.2cm} x^\bw_{jk} + \mathds{1}_{j=w_t} d^u_\bw, \notag \\[-3pt]
        & \hspace{5cm} \forall \bw \in \scrW, j \in \scrV, \label{eq:joint-cnst-demand}\\
        & \hspace{1cm}   \sum\limits_{i:(i,j) \in \scrA_\mathrm{R}} \hspace{-0.2cm} \big( x^r_{ij} + x^u_{ij} \big) = \hspace{-0.2cm} \sum\limits_{k:(j,k) \in \scrA_\mathrm{R}} \hspace{-0.2cm} \big( x^r_{jk} + x^u_{jk} \big), \label{eq:joint-cnst-rebalance}\\[-10pt]
        & \hspace{6cm}  \forall j \in \scrV_\mathrm{R},  \notag\\
        & \hspace{1.4cm} \bx^\bw_{ij} \geq 0, \quad \forall \bw \in \scrW, (i,j) \in \scrA, \label{eq:joint-cnst-nonnegative1} \\
        & \hspace{1.4cm} \bx^r_{ij} \geq 0, \quad \forall \bw \in \scrW,  (i,j) \in \scrA_\mathrm{R}, \label{eq:joint-cnst-nonnegative2} 
    \end{flalign}
\end{subequations}
where we use bold notation $\bx$ to represent a vector containing all the elements of $x_{ij}$. Moreover,  constraints \eqref{eq:joint-cnst-demand} take care of flow conservation and demand compliance as in a multi-commodity transportation problem (including flow conservation on the walking network), constraints \eqref{eq:joint-cnst-rebalance} ensure the rebalancing of the AMoD fleet (only on the road network), and the last two sets of constraints \eqref{eq:joint-cnst-nonnegative1}-\eqref{eq:joint-cnst-nonnegative2} restrict the flows to non-negative values. By solving \eqref{eq:problem-formulation} we find the optimal AMoD user and rebalancing flows. Note that the AMoD users' flow may consist of both walking or vehicle options, whereas the rebalancing flow is only for AMoD vehicles.

The objective $J$ is composed of two terms. 
The first term considers the total travel time of AMoD users. This term evaluates the travel time function $t_{ij}(x_{ij})$ with respect to the total flow (see \eqref{eq:total-flow}) which includes variables corresponding to private vehicle flow $x^p_{ij}$ (assumed to be fixed), and the rebalancing flow $x^r_{ij}$. Hence, when taking the product of $t_{ij}(x_{ij}) x^u_{ij}$ we obtain a non-convex function.
To address the non-convexity issue, we use a piecewise-affine approximation of $t_{ij}(x_{ij})$ which is further presented in Section \ref{sec:travel-latency-function-approximation}.  
The second term, i.e., $\bc'\bx^r$,  acts as a linear reguralizer whose purpose is to penalize rebalancing flows. This will ensure that a cost for rebalancing of the fleet is taken into account. 
In this work, we use $\bc = \lambda \bt^0$.
One can think of this reguralizer as a linear travel time function with respect to the rebalancing flow (since $(\lambda \bt^0)'\bx^r$). Therefore, if one lets $\lambda$ be high, with respect to the overall travel time, the reguralizer term will dominate the objective. Hence, we use a small $\lambda$ in order to guide the rebalancing flow through good paths without dominating the AMoD user routing decisions.

\subsection{Private Vehicle Flow Modeling} \label{subsec:private-vehicle-modeling}

Aiming to understand the interaction between a system-centric AMoD fleet and self-interested private vehicles, we assume some rationale behind private vehicle decisions.
To model this class of vehicles we use the \emph{user-centric} approach as in the Traffic Assignment Problem (TAP)  \cite{patriksson2015traffic}. This model finds, given OD demands, the flows in the network which achieve a Wardrop equilibrium \cite{wardrop1952road}.

Given a demand $\bg^p$ for these type of vehicle, each private user decides its route such that it minimizes its own travel time. Moreover, we impose that private vehicles can travel exclusively through the road network $\scrG_\mathrm{R}$. In other words, we do not allow private vehicles to change their transportation mode to walking.

Let $x^{p,\bw}_{ij}$ be the flow on link $(i,j)$ induced by private vehicle demand $d^p_{\bw}$ of OD pair $\bw$. Then, we assume private vehicles decide their routes by using the user-centric approach, 
\begin{subequations}\label{eq:TAP}
     \begin{flalign}
        &\min_{\bx^{p}} \hspace{0.3cm} \sum\limits_{(i,j) \in \scrA_\mathrm{R}}  \hspace{0.2cm} {\int\limits_{x^u_{ij}+x^r_{ij}}^{x_{ij}}{t_{ij}(s)ds}}  \label{eqn: TAP objective} \\
        &\text{s.t}  \sum\limits_{i:(i,j) \in \scrA_\mathrm{R}} \hspace{-0.2cm} x^{p, \bw}_{ij} + d^p_\bw \mathds{1}_{j=w_s} = \hspace{-0.5cm} \sum\limits_{k:(j,k) \in \scrA_\mathrm{R}} \hspace{-0.2cm} x^{p, \bw}_{jk} + d^p_\bw \mathds{1}_{j=w_t} , \notag \\
        & \vspace{-1.5cm} \hspace{4cm} \quad \forall \bw \in \scrW, j \in \scrV_\mathrm{R}, \label{eq:TAP-cnst-demand}\\ 
        & \quad\quad\quad \bx^{p, \bw} \geq \textbf{0}. \label{eq:TAP-cnst-nonnegativa}
    \end{flalign}
\end{subequations}
Notice that this version of the user-centric TAP is slightly different from the typical one~\cite{patriksson2015traffic}, given that it considers the AMoD flow in its objective (see limits of the integral on \eqref{eqn: TAP objective}).

To solve this problem we assume that the AMoD flow is fixed and private vehicles plan their routes considering AMoD flows as exogenous. When using this restriction, we can use the \emph{Method of Successive Averages} (MSA) \cite{daganzo1977stochastic} to solve \eqref{eq:TAP}.
Let us use the shorthand notation of $\texttt{TAP}(\bg, \bx^e)$ to indicate the TAP with $\bx^e$ being the exogenous flow. We denote a solution to  \eqref{eq:TAP} by  $\bx^p = \min \texttt{TAP}(\bg^p, \bx^u+\bx^r)$.

\subsection{Nested Problem for AMoD in Mixed Traffic} \label{subsec:bilevel-model}
Critically, AMoD flows react to the decisions made by private vehicles and these, in turn, react to private vehicles' flows. Hence, whenever private vehicles make their routing decisions, the AMoD fleet adjusts theirs, and vice versa.
This creates a nested optimization problem between these two classes of vehicles.
To give a formal definition of this game-theoretical problem we use the following bi-level optimization problem formulation
\begin{subequations} \label{eq:bilevel-problem-formulation}
    \begin{flalign} 
        & \hspace{-.3cm} \min\limits_{\{\bx^{\bw}\}_{\bw \in \scrW} ,\bx^r, \bx^p} J(\bx)  \\[3pt]
        & \hspace{.5cm}\text{s.t.} \hspace{0.3cm}   \eqref{eq:joint-cnst-demand}-\eqref{eq:joint-cnst-nonnegative2},  \\
        & \hspace{1.4cm} \bx^p \in \arg\min \texttt{TAP}(\bg^p, \bx^u+\bx^r), \label{eq:bilevel-cnst-TAP} 
    \end{flalign}
\end{subequations}
which has the same structure as~\eqref{eq:problem-formulation} with the additional constraint~\eqref{eq:bilevel-cnst-TAP}. The latter constraint refers to the TAP (the lower-level problem), which depends on the solution of the full problem (upper-level). Note that the upper-level problem is minimizing over the AMoD users, rebalancing, and privately-owned vehicle flows.

This phenomenon has been identified and is often described in a \emph{Stackelberg game} framework. In this setting, there is a \emph{leader} agent (in our case the AMoD manager) and a \emph{follower} (the private vehicles). In transportation networks, Korilis et al. \cite{korilis1997achieving} derived sufficient conditions to solve this problem when the network has parallel links. Under a similar setting, Lazar et al.~\cite{lazar2017capacity} have analyzed the links' capacity and price of anarchy for mixed traffic. Although these models enable a better understanding of the phenomenon, they are not applicable to general networks and one can hardly assess the benefits of system-centric routing in realistic networks.
To address this limitation, we will leverage the iterative approach~\cite{houshmand2019penetration} to compute an equilibrium between the private vehicles' and AMoD flows.

\subsection{Discussion}
A few comments are in order.
First, we assume the demand to be time-invariant. This assumption is in line with densely populated urban environments, where requests change slower compared to the average duration of a trip.
Second, we use the BPR function to relate traffic flows to travel time and allow flows to be fractional. While not capturing microscopic traffic phenomena, these approximations stem from established modeling assumptions suiting the mesoscopic perspective of our study.
    \section{AMoD Routing and Rebalancing Problem} \label{sec:travel-latency-function-approximation}

As mentioned earlier, the problem of routing and rebalancing as stated in \eqref{eq:problem-formulation} is non-convex for typical travel time functions such as BPR. This happens due to the term $t(x_{ij})x^r_{ij}$ in the objective function which takes products of the form $k  (x^u_{ij})^n x^r_{ij}$ with $k$ and $n$ being a constant and the order of the polynomial, respectively. To overcome this issue, we take the suggested piecewise-affine approximation in~\cite{SalazarTsaoEtAl2019} and  extend it to a 3-lines approximation.
We first present the analysis for the 2-line Congestion-Aware Routing Scheme (CARS)~\cite{SalazarTsaoEtAl2019} and then extend it to the 3-lines segment case (CARS3). Finally, we present a disjoint formulation of the problem which will serve as a benchmark for comparison. 

\subsection{2-line Piecewise-affine Approximation (CARS)}\label{subsec:2-line-pwlinear-approx}

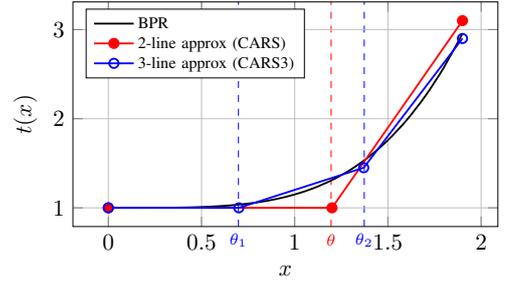
\begin{figure}
    \centering
    \begin{tikzpicture}
    \draw [dashed, color=red] (3.9,0) -- (3.9, 3.42);
    
    
    \draw [color=red] (3.9,-0.2) node{{\scriptsize $\theta$}};
    
    \draw [dashed, color=blue] (2.5,0) -- (2.5, 3.42);
    
    \draw [color=blue] (2.5,-0.2) node{{\scriptsize $\theta_1$}};
    
    
    \draw [dashed, color=blue] (4.4,0) -- (4.4, 3.42);
    
    
    \draw [color=blue] (4.4,-0.2) node{{\scriptsize $\theta_2$}};
    
    \begin{axis}[
        xlabel=$x$,
        ylabel=$t(x)$,
        ylabel near ticks,
        width = 8cm,
        height = 5cm,
        legend pos= north west,
        legend cell align={left},
        legend style={nodes={scale=0.7, transform shape}},
        grid=both,
        grid style={line width=.1pt, draw=gray!10},
        major grid style={line width=.2pt,draw=gray!50}
        ]
    \addplot[smooth,black, style={thick}] plot coordinates {
        ( 0, 1 )
        ( 0.1, 1.000015 )
        ( 0.2, 1.00024 )
        ( 0.3, 1.001215 )
        ( 0.4, 1.00384 )
        ( 0.5, 1.009375 )
        ( 0.6, 1.01944 )
        ( 0.7, 1.036015 )
        ( 0.8, 1.06144 )
        ( 0.9, 1.098415 )
        ( 1, 1.15 )
        ( 1.1, 1.219615 )
        ( 1.2, 1.31104 )
        ( 1.3, 1.428415 )
        ( 1.4, 1.57624 )
        ( 1.5, 1.759375 )
        ( 1.6, 1.98304 )
        ( 1.7, 2.252815 )
        ( 1.8, 2.57464 )
        ( 1.9, 2.954815 )
    };
    \addlegendentry{BPR}

    \addplot[color=red, mark=*, style={thick}] plot coordinates {
        (0,1)
        (1.2, 1)
        (1.9, 3.1)
    };
    \addlegendentry{2-line approx (CARS)}

    \addplot[color=blue, mark=o, style={thick}] plot coordinates {
        (0,1)
        (0.7, 1)
        (1.37, 1.45)
        (1.9, 2.9)
    };
    \addlegendentry{3-line approx (CARS3)}
    \end{axis}

    \end{tikzpicture}
    \vspace{-.2cm}
    \caption{Travel time function approximation}
    \vspace{-.5cm}
    \label{fig:BPR-approximation}
\end{figure}

Recall that the non-convexity in~\eqref{eq:joint-obj} arises from the product of the AMoD users' flow $x^u_{ij}$ with the rebalancing flow $x^r_{ij}$. Hence, we aim to approximate this term with a convex function which makes it more computationally efficient, and therefore gives tractability to larger instances of the problem. Specifically, we approximate the latency function (Eq. \eqref{eq:latency-function}) using a piecewise-affine function as shown in Fig.~\ref{fig:BPR-approximation}. Let such a function be 
\begin{alignat}{2} \label{eq:2-line-pwlinear}
\hat{t}_{ij}(x) = & 
    \begin{dcases}
        at_{ij}^0, & \text{if } x<\theta_{ij}, \\
        at_{ij}^0 + b_{ij}(\theta_{ij}-x), & \text{if } x \geq \theta_{ij},\\
    \end{dcases}
\end{alignat}
where $a$ and $b_{ij}$ are constant values. In our case, we assume $a=1$; and $b_{ij} = \beta/m_{ij}$ with $\beta$ being the slope of the second segment. Let the non-smooth threshold of the function be $\theta_{ij} = m_{ij}\theta$, where $\theta$ is the threshold in the normalized travel time function. 
In order to model this non-smooth function in the optimization problem, we introduce the set of slack variables $\varepsilon_{ij}$ defined as 
\begin{equation} \label{eq:2-line-epsilon}
   \varepsilon_{ij} = \max \{0, x_{ij}-\theta_{ij} \} ,
\end{equation}
which denotes the exceeding flow after threshold $\theta_{ij}$. In the optimization problem \eqref{eq:problem-formulation} we model these variables by adding linear constraints $\varepsilon_{ij} \geq 0$ and $\varepsilon_{ij} \geq \theta_{ij}-x$, provided that the objective is a function of $\varepsilon_{ij}$. With these definitions we are ready to analyze and propose a tractable cost function. To this end, we focus attention on an element-wise analysis of the first term (non-convex part) of objective \eqref{eq:joint-obj} using $\hat{t}$ instead of $t$, which we call $\hat{J}_{ij}$. 
\begin{subequations}
    \begin{flalign}
        \hat{J}_{ij} &=  \hat{t}_{ij}( x_{ij} ) x^u_{ij} \label{eq:j-hat-orignal}\\
        & =  t^0_{ij}(a+b_{ij}\varepsilon_{ij})x^u_{ij} \\
        & =  at^0_{ij}x^u_{ij} + t^0_{ij}b_{ij}\varepsilon_{ij}x^u_{ij} \\
        & =  at^0_{ij}x^u_{ij} + b_{ij}t^0_{ij}\varepsilon_{ij}(\varepsilon_{ij}+\theta_{ij}-x^r_{ij}-x^p_{ij}) \label{eq:j-derivation-eps}\\
        & \leq  at^0_{ij}x^u_{ij} + b_{ij}t^0_{ij}\varepsilon_{ij}(\varepsilon_{ij}+\theta_{ij}-x^p_{ij}) \label{eq:j-derivation-final-step}
    \end{flalign}
\end{subequations}
where in \eqref{eq:j-derivation-eps} we express $x^u_{ij}$ by using a combination of \eqref{eq:total-flow} and \eqref{eq:2-line-epsilon}; in the last step \eqref{eq:j-derivation-final-step}, we add to $\hat{J}_{ij}$ the term $b_{ij}t^0_{ij}\varepsilon_{ij}x^r_{ij}$. 
By adding this term to $\hat{J}$, we consider a relaxation of the original problem (i.e., minimizing an upper bound of $\hat{J}$ \eqref{eq:j-derivation-final-step} as opposed to the original $\hat{J}$ in \eqref{eq:j-hat-orignal}). This relaxation allows the proposed objective to be quadratic. Let the relaxed objective be
\begin{subequations}
    \begin{align}
        J^{\text{QP}}_{ij} &= at^0_{ij}x^u_{ij} + b_{ij}t^0_{ij}\varepsilon_{ij}(\varepsilon_{ij}+\theta_{ij}-x^p_{ij}) \\
        &= \hat{t}_{ij}( x_{ij} ) x^u_{ij} + \hat{t}^{a=0}_{ij}( x_{ij} )x^r_{ij}.
    \end{align}
\end{subequations}
where $\hat{t}^{a=0}(x)$ is equal to $\hat{t}(x)$ with parameter $a=0$. 
By analyzing this convex approximation  $J^{\text{QP}}$ with both $J$ and $\hat{J}$, we observe that the implication of adding the extra term is taking into account congestion-aware rebalancing when the flow is greater than $\theta_{ij}$. Nevertheless, this congestion-aware routing of the rebalancing vehicles has a lower impact in $J^{\text{QP}}$ than the AMoD users flows since $a=0$ in $\hat{t}^{a=0}_{ij}(x_{ij})x^r_{ij}$, i.e., the function starts to increase from an initial point equal to zero instead of $t^0_{ij}$.

Considering that the number of rebalancing vehicles has a minor impact on $J$ in comparison to road congestion,  and the fact that it converges to zero for perfectly symmetric demand distributions \cite{RossiZhangEtAl2017}, $J^{\text{QP}}$ can be used as a model to estimate the total travel time on road arcs. Our empirical studies show that, when no rebalancing is considered, the difference between the solution $J^*$ and $J$ evaluated with the optimal solution of the Quadratic Program (QP) is typically less than $5\%$ (Figs.  \ref{fig:EMA-precision} and \ref{fig:NYC-precision}).

\subsection{3-line Piecewise-affine Approximation (CARS3)} \label{subsec:3-line-pwlinear-approx}

Given that CARS might not provide a very accurate estimate of travel times when the flow is around the capacity level (Fig. \ref{fig:BPR-approximation}), we next approximate the travel time function using a more accurate 3-line piecewise-affine function. To construct this function, we will follow the same analysis as in the 2-lines case. The price to pay for increasing the precision of the function is that it requires adding $|\scrA|$ (number of arcs) extra variables and $|\scrA|$ new linear constraints to the optimization problem.
Following the same analysis as in the previous section we define
\begin{subequations}
    \begin{alignat*}{3}
    \hat{t}_{ij}(x) = & 
        \begin{dcases}
            at_{ij}^0, \ \quad \quad \quad \quad \quad \quad \quad \quad \quad \quad \text{if } x<\theta_{ij} \\
            at_{ij}^0 \hspace{-.1cm}+\hspace{-.1cm} b_{ij}(\theta^{(1)}_{ij}\hspace{-.1cm}-\hspace{-.1cm}x), \quad\quad\quad\quad\quad \text{if } \theta^{(1)}_{ij} \leq x \leq \theta^{(2)}_{ij} \\
            at_{ij}^0 \hspace{-.1cm}+\hspace{-.1cm} b_{ij}(\theta^{(2)}_{ij}\hspace{-.1cm}-\hspace{-.1cm}\theta^{(1)}_{ij})\hspace{-.1cm} + \hspace{-.1cm}c_{ij}(\theta^{(2)}_{ij} \hspace{-.1cm}-\hspace{-.1cm}x), \quad \text{if } \theta^{(2)}_{ij} \leq x,\\
        \end{dcases}
    \end{alignat*}
\end{subequations}
where $a$, $b_{ij}$ and $c_{ij}$ are constant values with $a=1$; $b_{ij} = \beta/m_{ij}$; and $c_{ij} = \sigma/m_{ij}$. The slope of the function is $\beta$ for $x_{ij}\in(\theta^{(1)}_{ij}, \theta^{(2)}_{ij})$ and $\sigma$ for $x_{ij}>\theta^{(2)}_{ij}$. Moreover,  $\theta^{(1)}$ and $\theta^{(2)}$ are the normalized, non-smooth thresholds of the travel time function. 
Assuming $\theta^{(2)}_{ij} \geq \theta^{(1)}_{ij}$  and $\sigma, \beta > 0 $ we define two new sets of slack variables as
\begin{subequations}
    \begin{align}
        \varepsilon^{(1)}_{ij} &= \max\{0, x_{ij} - \theta^{(1)}_{ij} - \varepsilon^{(2)}_{ij}\},  \\
        \varepsilon^{(2)}_{ij} &= \max\{0, x_{ij} - \theta^{(2)}_{ij}\},
    \end{align}
\end{subequations}
where $\varepsilon^{(1)}_{ij}$ is the excess flow after $\theta^{(1)}_{ij}$ and up to $\theta^{(2)}_{ij}-\theta^{(1)}_{ij}$, and $\varepsilon^{(2)}_{ij}$ is the excess flow after $\theta^{(2)}_{ij}$. Note that $ \varepsilon^{(1)}_{ij}$ is defined in terms of $\varepsilon^{(2)}_{ij}$  to ensure that it is upper-bounded by $\theta^{(2)}_{ij}-\theta^{(1)}_{ij}$. Using the same analysis as in the 2-lines case, we get
\begin{subequations}
    \begin{flalign}
        \hat{J}_{ij} &= \hat{t}_{ij}( x_{ij} ) x^u_{ij} \\
        &= at^0_{ij}x^u_{ij} + b_{ij}t^0_{ij}\varepsilon^{(1)}_{ij}(\varepsilon^{(1)}_{ij}+\varepsilon^{(2)}_{ij}+\theta^{(1)}_{ij}-x^r_{ij}-x^e_{ij}) \notag \\
        & \quad \quad + c_{ij}t^0_{ij}\varepsilon^{(2)}_{ij}(\varepsilon^{(2)}_{ij}+\theta^{(2)}_{ij}-x^r_{ij}-x^e_{ij})  \\
        & \leq at^0_{ij}x^u_{ij} + b_{ij}t^0_{ij}\varepsilon^{(1)}_{ij}(\varepsilon^{(1)}_{ij}+\varepsilon^{(2)}_{ij}+\theta^{(1)}_{ij}-x^e_{ij}) \notag \\
        & \quad \quad + c_{ij}t^0_{ij}\varepsilon^{(2)}_{ij}(\varepsilon^{(2)}_{ij}+\theta^{(2)}_{ij}-x^e_{ij}) \label{eq:3-line-derivations}.
    \end{flalign}
\end{subequations}
We add the rebalancing variables as in the CARS to get \eqref{eq:3-line-derivations}.  Even though the term $b_{ij}t^0_{ij}\varepsilon^{(1)}_{ij}\varepsilon^{(2)}_{ij}$ is not guaranteed to be convex, $\varepsilon^{(1)}_{ij}\varepsilon^{(2)}_{ij}=0$  if $x_{ij}<\theta^{(2)}_{ij}$. Additionally, notice that when $x_{ij}>\theta^{(2)}_{ij}$ the residual flow $\varepsilon^{(1)}_{ij}=(\theta^{(2)}_{ij}-\theta^{(1)}_{ij})$. Therefore, we can replace $b_{ij}t^0_{ij}\varepsilon^{(1)}_{ij}\varepsilon^{(2)}_{ij}$ with $b_{ij}t^0_{ij}(\theta^{(2)}_{ij}-\theta^{(1)}_{ij})\varepsilon^{(2)}_{ij}$ and write the objective function of the QP as
\begin{flalign}
    J^{\text{QP}}_{ij} &= at^0_{ij}x^u_{ij} + b_{ij}t^0_{ij}\varepsilon^{(1)}_{ij}(\varepsilon^{(1)}_{ij}+\theta^{(1)}_{ij}-x^e_{ij})  \notag \\
    & \quad\quad + c_{ij}t^0_{ij}\varepsilon^{(2)}_{ij}(\varepsilon^{(2)}_{ij}+\theta^{(2)}_{ij}-x^e_{ij}) \label{eq:QP-objective-CARS3} \\
    & \quad\quad+ b_{ij}t^0_{ij}(\theta^{(2)}_{ij}-\theta^{(1)}_{ij})\varepsilon^{(2)}_{ij} \notag \\
    &= \hat{t}_{ij}( x_{ij} ) x^u_{ij} + \hat{t}^{a=0}_{ij}( x_{ij} )x^r_{ij}, \notag
\end{flalign}
where $\varepsilon^{(1)}_{ij}$ and $\varepsilon^{(2)}_{ij}$ are linearly constrained as follows
\begin{flalign}
    &\varepsilon^{(1)}_{ij} \geq 0, \quad 
    \varepsilon^{(1)}_{ij} \geq x_{ij} - \theta^{(1)}_{ij} - \varepsilon^{(2)}_{ij}, \label{eq:epsilon-cnstr2}\\
    &\varepsilon^{(2)}_{ij} \geq 0,  \quad
    \varepsilon^{(2)}_{ij} \geq x_{ij} - \theta^{(2)}_{ij}. \label{eq:epsilon-cnstr4}
\end{flalign}

As a result, we get a better convex approximation of the original problem compared to CARS model. The QP problem is then to minimize \eqref{eq:QP-objective-CARS3} subject to \eqref{eq:joint-cnst-demand}-\eqref{eq:joint-cnst-nonnegative2}, and \eqref{eq:epsilon-cnstr2}-\eqref{eq:epsilon-cnstr4}.

An important trade-off worth noting is the difference between CARS and CARS3. Even though CARS3 provides a better approximation of the cost function and hence a better solution to the problem, it requires $|\scrA|$ additional variables and constraints.

\subsection{Disjoint Strategy} \label{subsec:disjoint}

Another way of addressing the system-centric routing and re-balancing problem is to solve the problem using a disjoint method instead of the joint approach.
That is, to solve the system-centric problem for AMoD users first, and then solve the rebalancing problem formulated as a linear program (LP).
A formal definition of this problem is first solving
\begin{flalign} 
    & \min\limits_{\{\bx^{\bw}\}_{\bw \in \scrW}}   \sum\limits_{(i,j)\in \scrA} t_{ij}( x_{ij} ) x^u_{ij} ,    
    \quad  \text{s.t.} \quad  \eqref{eq:joint-cnst-demand}, \eqref{eq:joint-cnst-nonnegative1} \label{eq:disjoint-routing},
\end{flalign}
and then, use the resulting optimal $\bx^{u*}$ as an input to 
\begin{flalign} 
    & \min\limits_{\bx^r}  \quad \bc' \bx^r
    , \quad \text{s.t.} \quad \eqref{eq:joint-cnst-rebalance}, \eqref{eq:joint-cnst-nonnegative2} \label{eq:disjoint-rebalace},
\end{flalign}
It is important to point out that the system-centric Problem~\eqref{eq:disjoint-routing} is a constrained nonlinear program (NLP) which might take time to solve.
In contrast to the disjoint formulation, the methodology we propose (CARS3) offers the possibility to solve the problem as a QP, which is usually faster than a higher order NLP and provides global optimality guarantees.

\subsection{Iterative Solution Nested Problem}\label{subsec:iterative}
\begin{figure}[t]
	\centering
	\begin{subfigure}{.49\linewidth}
		\begin{tikzpicture}[
        node distance= 9em and 5em,
        sloped]
        \tikzset{
        box/.style = {
            fill=black!8,
            shape=rectangle, 
            rounded corners,
                draw=black!40, 
                align=center,
                text = black,
                font=\fontsize{10}{10}\selectfont},
        dummybox/.style = {
            shape=circle,  
                align=center, 
                },
        arrow/.style={
            color=black,
            draw=black,
            -latex,
                font=\fontsize{8}{8}\selectfont},
        }
        \node[box](B1){Solve user-centric for \\private-vehicle \eqref{eq:TAP}};
        \node[box, below=1cm of B1](S1){\ Solve system-centric \\ for AMoDs \eqref{eq:problem-formulation}};
        \draw[arrow](S1.west) to [out=190,in=180] node[below]{Fix AMoD flow} (B1.west);
        
        \draw[arrow](B1.east) to [out=10,in=-0] node[below]{Fix private-vehicle flow} (S1.east);
        
\end{tikzpicture}
		\caption{}
		\label{fig:iterative-sketch}
	\end{subfigure}
	\begin{subfigure}{.49\linewidth}
		\includegraphics[width=0.85\linewidth]{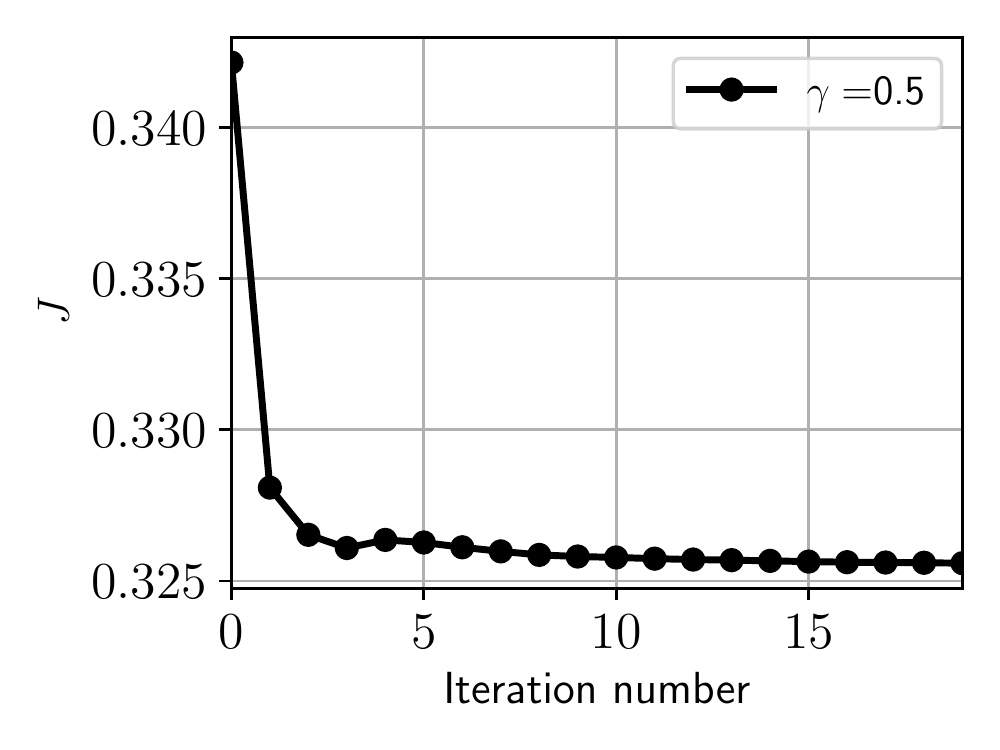}
		\vspace{-.2cm}
		\caption{}
		\label{fig:iterative-convergance}
	\end{subfigure}
	\caption{(a): A sketch of the procedure for solving the bi-level problem \eqref{eq:bilevel-problem-formulation}. (b): An example of the total cost converging for an AMoD penetration rate of $0.5$ on the NYC sub-network}
	\label{fig:iterative-algorithm-pic}
	\vspace{-.6cm}
\end{figure}
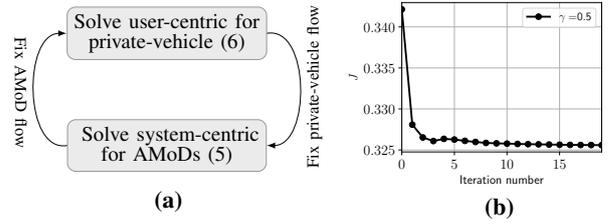

To compute an equilibrium for the nested problem~\eqref{eq:bilevel-problem-formulation} outlined in Section~\ref{subsec:bilevel-model}, we use the framework developed in~\cite{houshmand2019penetration} which uses an iterative approach to reach an equilibrium between the private and AMoD flows (Fig.~\ref{fig:iterative-algorithm-pic}).
Instead of solving the bi-level Problem~\eqref{eq:bilevel-problem-formulation}, we solve Problem~\eqref{eq:problem-formulation} with one of the methods presented in this Section (CARS, CARS3 or Disjoint) and \eqref{eq:TAP} iteratively and use the output of each problem as the input to the other one. 
In other words, consider a private vehicle demand $\bg^{u}$ and solve $\bx^p=\min \texttt{TAP}(\bg^p, \mathbf{0})$. Then, solve the AMoD routing and rebalancing problem~\eqref{eq:problem-formulation} for AMoD demand $\bg^u$ with fixed input $\bx^p$ (the solution of the previously solved TAP).
Since private vehicles were unaware of AMoDs in the system while solving the TAP, we solve again the problem considering a fixed flow equal to $\bx^u+\bx^r$, i.e., $\bx^p=\texttt{TAP}(\bg^p, \bx^u+\bx^r)$, and iterate this process until it converges as shown in Fig.~\ref{fig:iterative-convergance}.

Also, note that both the disjoint problem in Sec.~\ref{subsec:disjoint}, and the iterative model allow for updating the component $\bt^0$ in $\bc$ for the travel times $\bt(\bx)$ from the solution of  \eqref{eq:disjoint-routing} or previous iteration of the iterative method. This results in a more accurate cost function in terms of the travel time weight for the rebalancing problem.

We do not provide theoretical arguments on the uniqueness or stability of the players (AMoD and private vehicles) equilibria, due to the non-separability of the cost functions with respect to their individual players' strategies~\cite{harker1988multiple}. Yet, empirically, this iterative algorithm always converged in a few iterations to results that are consistent for different penetration rates. We leave the theoretical study of the properties of the equilibria found to future work.

    \section{Experiments} \label{sec:experiments}

\begin{figure}[t]
    \centering
    \begin{subfigure}{.49\linewidth}
      \centering
      \includegraphics[trim={0 0cm 0 0},clip, width=1\linewidth]{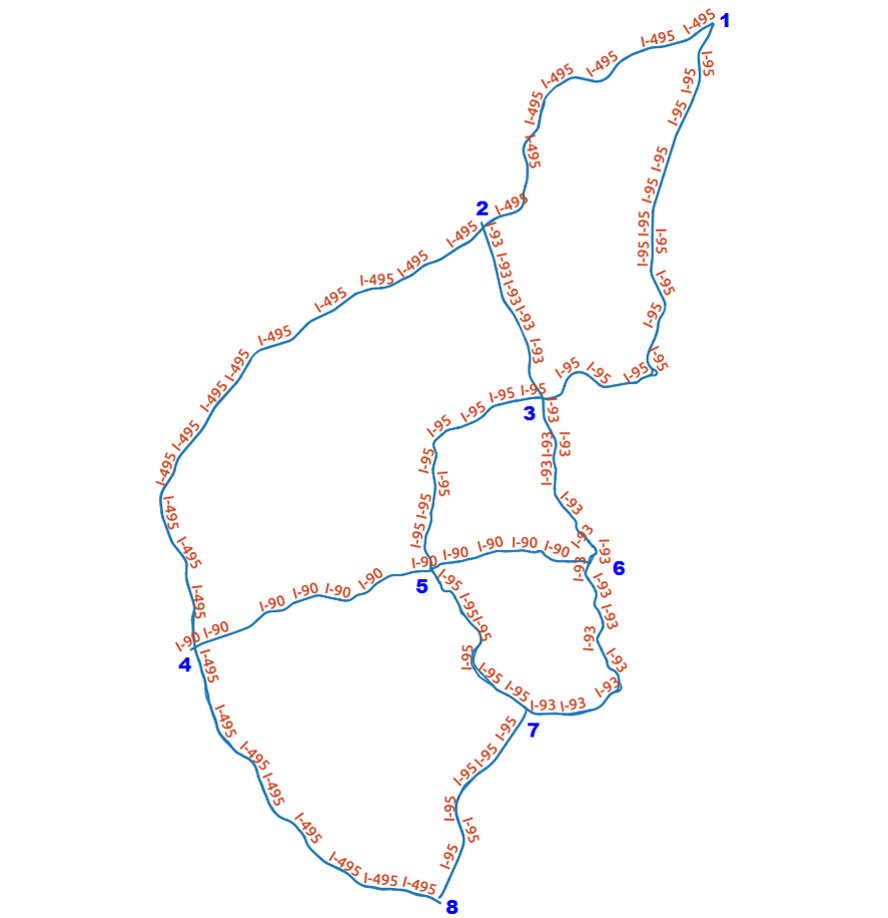}
      \caption{EMA subnetwork}
      \label{fig:EMA-net}
    \end{subfigure}
    \begin{subfigure}{.49\linewidth}
      \centering
      \includegraphics[trim={0cm 0cm 2.5cm 0},clip, width=0.39\linewidth]{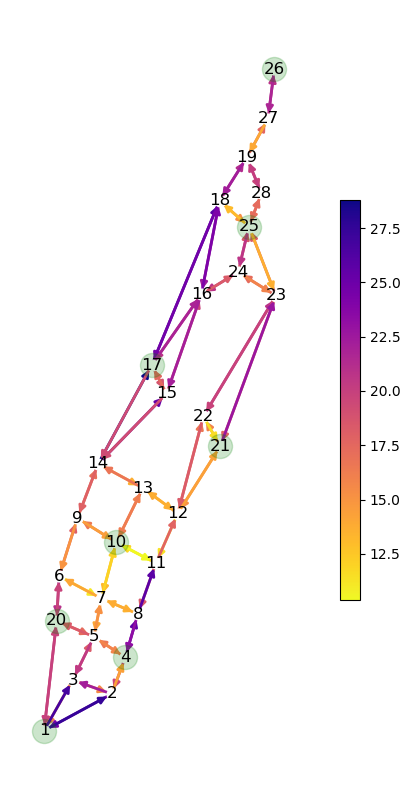}
      \caption{NYC subnetwork}
      \label{fig:NYC_net}
    \end{subfigure}
    \caption{Subnetworks used for the experiments.}
    \vspace{-.6cm}
\end{figure}

 In order to validate our proposed routing algorithms, we consider two data-driven case studies on sub-networks of Eastern Massachusetts (EMA) interstate highways and New York City (NYC). The EMA road network (Fig. \ref{fig:EMA-net}) consists of $8$ nodes and $24$ links. We consider every node as a zone (origin-destination candidate) which results in $56$ OD pairs. The NYC network was built using two data sources: OpenStreetMaps \cite{OpenStreetMap} from which we retrieve the network topology and road characteristics, and the recently released \emph{Uber Movement Speed Data set} \cite{UberData} which was used to assign speed data to road segments (available hourly). We build a sub-network  (Fig.~\ref{fig:NYC_net}) consisting of $28$ nodes, $90$ edges and $8$ zones (green dots).

We use the three methodologies described in Sec. \ref{sec:travel-latency-function-approximation} (CARS, CARS3 and Disjoint) to solve the fleet routing and rebalancing problem and compare their results against each other. Our first two experiments reveal that using CARS and CARS3 result in accurate solutions with low running times for these networks.

\subsection{Accuracy of CARS and CARS3}
Using numerical examples, we show how the optimal solution of CARS and CARS3 compare with the optimal solution of the system-centric problem. To achieve this, we consider the case in which rebalancing is not required, i.e., constraints \eqref{eq:joint-cnst-rebalance} are excluded and variables $\bx^r$ are set to zero. Then, the non-rebalanced routing problem becomes the system-centric traffic assignment problem with exogenous flow (problem \eqref{eq:disjoint-routing}). This problem is convex \cite{patriksson2015traffic} and can be solved using nonlinear programming (NLP) algorithms. 

\begin{figure}[t]
    \centering
    \begin{subfigure}{1\linewidth}
      \includegraphics[trim={0 1.22cm 0 0},clip, width=1\linewidth]{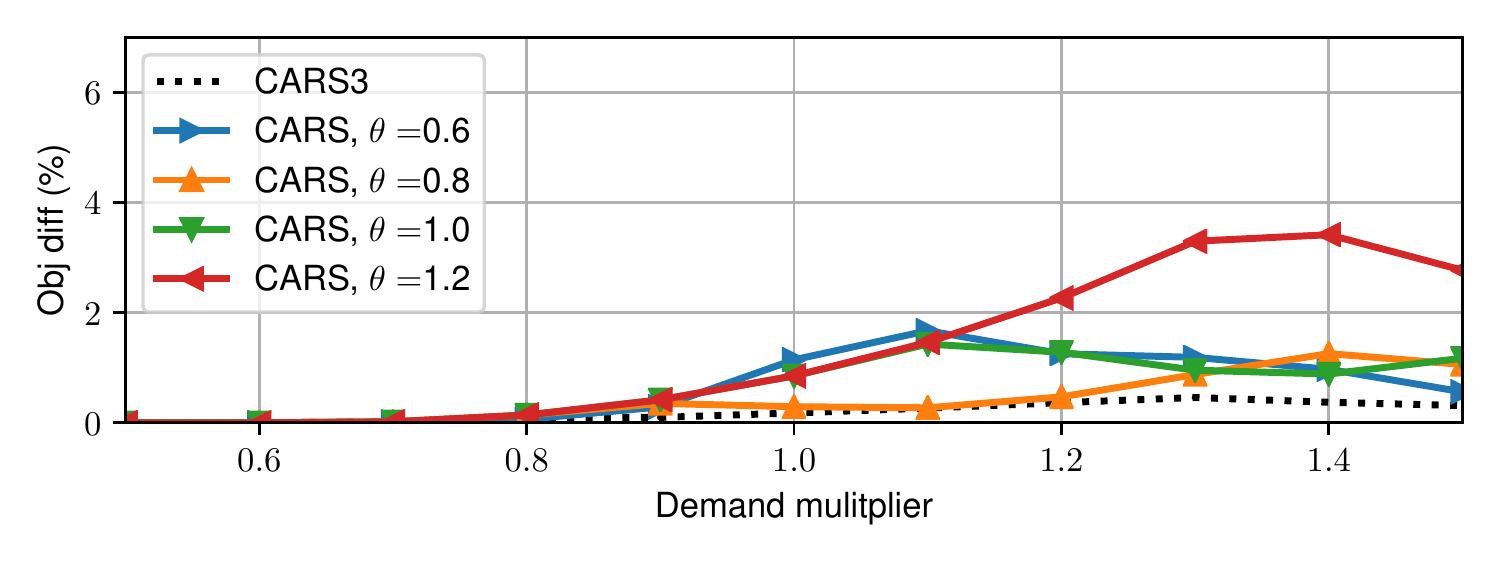}
      \caption{}
      \label{fig:EMA-precision}
    \end{subfigure}
    \begin{subfigure}{1\linewidth}
      \includegraphics[trim={0 0 0 0},clip, width=1\linewidth]{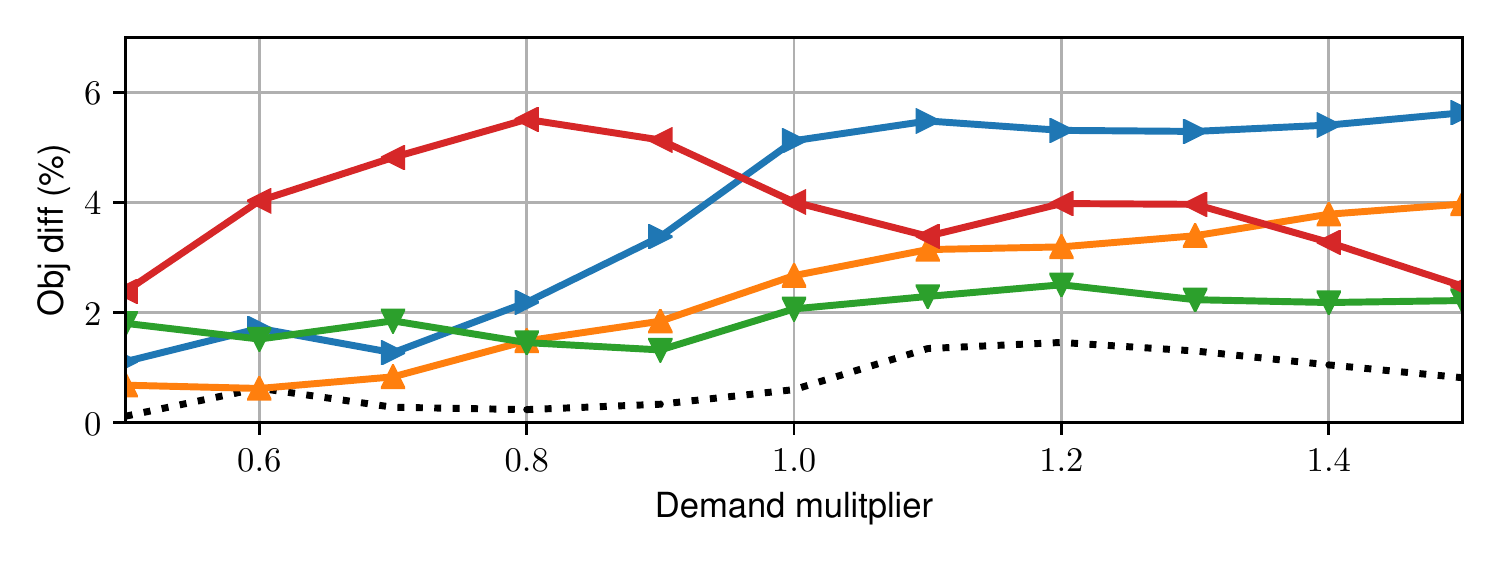}
      \vspace{-0.7cm}
      \caption{}
      \label{fig:NYC-precision}
    \end{subfigure}
    \caption{Deviation in percentage terms between the approximated model and the optimal solution of the non-rebalanced system-centric problem.  \textbf{(a)} and  \textbf{(b)} present results for the EMA and NYC networks, respectively.}
    \label{fig:precision}
\end{figure}

This experiment assesses the offset of the total cost between the approximate models (CARS, CARS3) and the optimal solution considering the non-rebalancing system-centric model. To make a fair comparison, the solution of CARS and CARS3 are evaluated in the original cost function $J(\bx)$ from~\eqref{eq:joint-obj}. We gather results for different traffic levels (demands) for the EMA (Fig.~\ref{fig:EMA-precision}) and NYC (Fig.~\ref{fig:NYC-precision}) networks.
The purpose of using different demands is to investigate the approximation quality of $\hat{t}(\cdot)$ (Fig.~\ref{fig:BPR-approximation}) at different flow levels. 
Note that for the two networks, the CARS3 model outperforms the CARS method for different thresholds $\theta$ and demand rates $\bg$. We attribute this behavior to the fact that the 3-lines model yields a more precise approximation to the travel time function than the 2-lines one. Moreover, we consider the CARS3 model to be a good approximation as its deviations are always below~2\%.

\subsection{Computational Time and Evaluation of the Cost}

We compare the running times of CARS, CARS3, and Disjoint as well as the quality of their solutions. For each approach, we solve $30$ problems by multiplying the OD demand vector $\bg$ by a uniform distributed random variable in the range of $[0.8, 1.2]$. For each run, we collect the computational time $\tau$ as well as $J$ which is computed by applying~\eqref{eq:joint-obj} to each solution. 
Table~\ref{table:computational-times} reports the mean $\mu_\tau$ and variance $\sigma_\tau$ of the computational time. Additionally, we report the average objective function divided by the total demand \mbox{$\bar{J} = J/(\sum_{\bw \in W} d_\bw)$}. 
\begin{table}[t]
    \centering
    \resizebox{\linewidth}{!}{%
    \begin{tabular}{|lc|ccc|ccc|}
    \hline
    \multirow{2}{*}{Model} & \multirow{2}{*}{Type} & \multicolumn{3}{c|}{EMA} & \multicolumn{3}{c|}{NYC} \\ \cline{3-8}
        & {} & $\mu_\tau$ [s] & $\sigma_\tau$ [s] & $\bar{J}$ & $\mu_\tau$ [s] & $\sigma_\tau$ [s] & $\bar{J}$ \\ \hline
    \textbf{CARS}& QP & \textbf{0.016} & 3e-7 & 0.427 &  \textbf{0.170} & 8e-4 & 0.324 \\
    \textbf{CARS3}& QP & \textbf{0.022} & 1e-6 & 0.422 & \textbf{0.215} & 3e-3 & 0.317 \\
    \textbf{Disjoint}& & \textbf{5.48} & 0.269 & 0.421 & \textbf{24.88} & 1.72 & 0.31 \\
    \  System-centric& NLP & \ 5.48 & 0.269 &  & \ 24.88 & 1.72 & \\
    \  Rebalance& LP & \ 2e-4 & 2e-10 & & \ 4e-5 & 2e-10  & \\ \hline
    \end{tabular}%
    }
    \caption{Computational times and objective function for different models and networks.  $\mu_\tau$ and $\sigma_{\tau}$ are the average computational time (seconds) and variance over 30 samples, respectively. The average cost is denoted with $\bar{J}$. }
    \label{table:computational-times}
    \vspace{-.6cm}
\end{table}
All the scenarios studied were performed using an Intel(R) Core(TM) \mbox{i7-8700} CPU @ 3.20GHz and 32 GB of RAM memory. To solve the NLP we used the IPOPT solver \cite{wachter2006implementation}, whereas the QP and LP programs were solved using Gurobi~8.1.1~\cite{gurobi}.

As expected, we observe that the disjoint model is the slowest, given that its first step requires solving a NLP (followed by a significantly faster solution of an LP). This method takes about $25$ and $100$ times more time than solving CARS3 for EMA and NYC, respectively. Moreover, given that CARS3 requires more variables and constraints, it takes around $30\%$ more time than CARS to solve. 

Furthermore, our results of $\bar{J}$ show that the Disjoint method finds the best solution between the three models. The reason for this is that its model for routing is not an approximation. Nevertheless, the solutions of CARS and CARS3 are less than $4\%$ and $2\%$ away from the Disjoint solution, respectively.
Arguably, this result might suggest that the benefit of solving the problem jointly is not as valuable as assumed, which coincides with the results of~\cite{RossiZhangEtAl2017}.
However, it is worth mentioning that these results are sensitive to different OD demand distributions. As an example, for perfectly symmetrical OD demands, rebalancing plays no role in the optimization process. 

\subsection{System-optimal Routing and Rebalancing Trade-off} \label{subsec:tradeoff-systemcentric-rebalancing}

Considering the existence of selfish privately-owned vehicles and centrally-controlled AMoD vehicles, we analyze the trade-off that exists between system-optimal AMoD routing and the additional traffic due to AMoD rebalancing in terms of average travel times. 
We tackle the bi-level Problem~\eqref{eq:bilevel-problem-formulation} following the iterative methodology presented in Section~\ref{subsec:iterative}. We use different penetration rates of AMoD customers with respect to the total demand. More specifically, we let $\gamma \in [0,1]$ be the penetration rate and  $\bg$ the total OD demand. Then, we assume that $\bg^u = \gamma \bg$ and $\bg^p = (1-\gamma)\bg$ are the AMoD's and private vehicles' demand, respectively. In this paper, we choose the same demand distribution for AMoD and private vehicles. Yet, different demand separation criteria can be readily implemented in this framework.
\begin{figure}[t] 
    \centering
    \begin{subfigure}{.515\linewidth}
      \centering
      \includegraphics[trim={0 0.9cm 0 0},clip, width=1\linewidth]{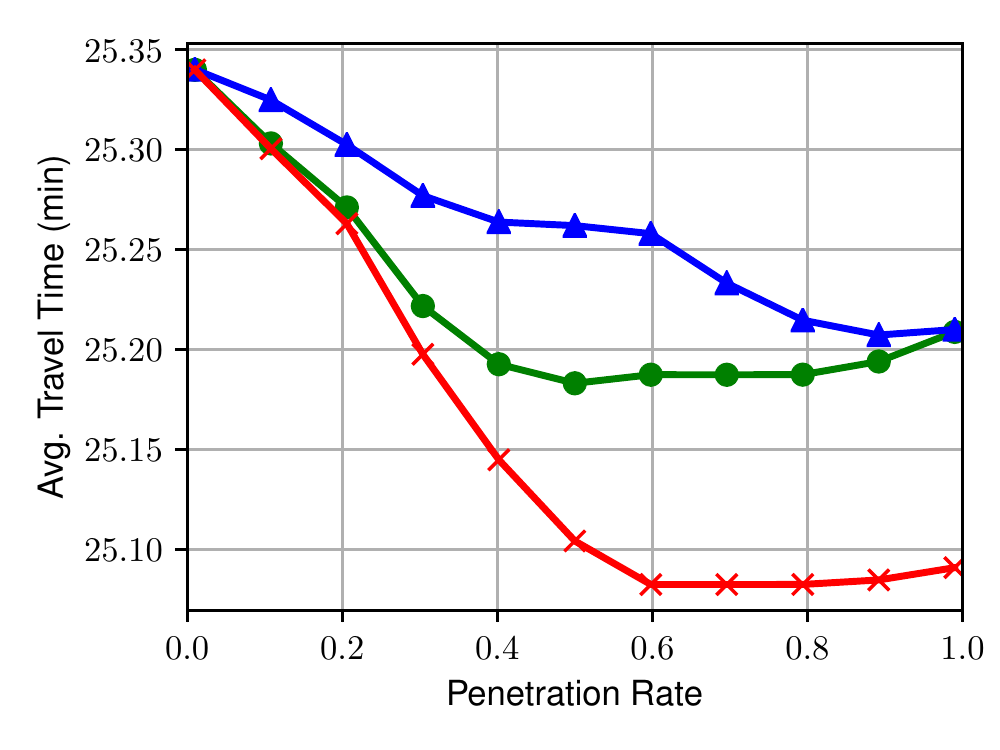}
      \caption{EMA, no rebalancing}
      \label{fig:EMA-penRate-no-rebalance}
    \end{subfigure}
    \begin{subfigure}{.465\linewidth}
      \centering
      \includegraphics[trim={0.8cm 0.9cm 0 0},clip, width=1\linewidth]{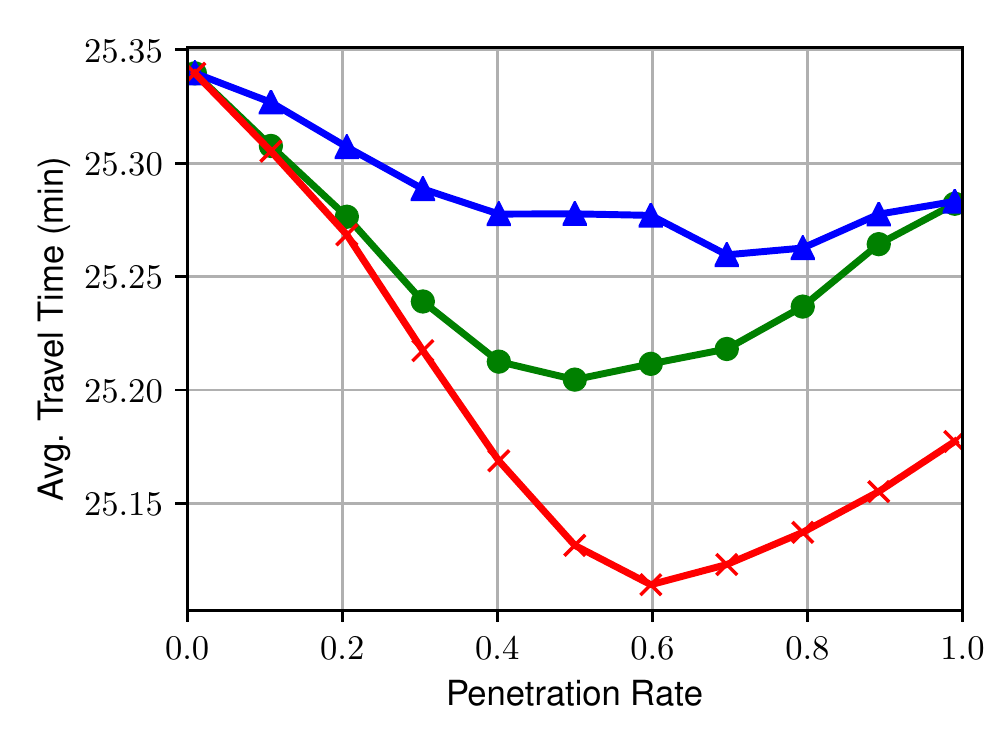}
      \caption{EMA, rebalancing}
      \label{fig:EMA-penRate-rebalance}
    \end{subfigure}
    \begin{subfigure}{.515\linewidth}
      \centering
      \includegraphics[trim={0 0.4CM 0 0},clip, width=1\linewidth]{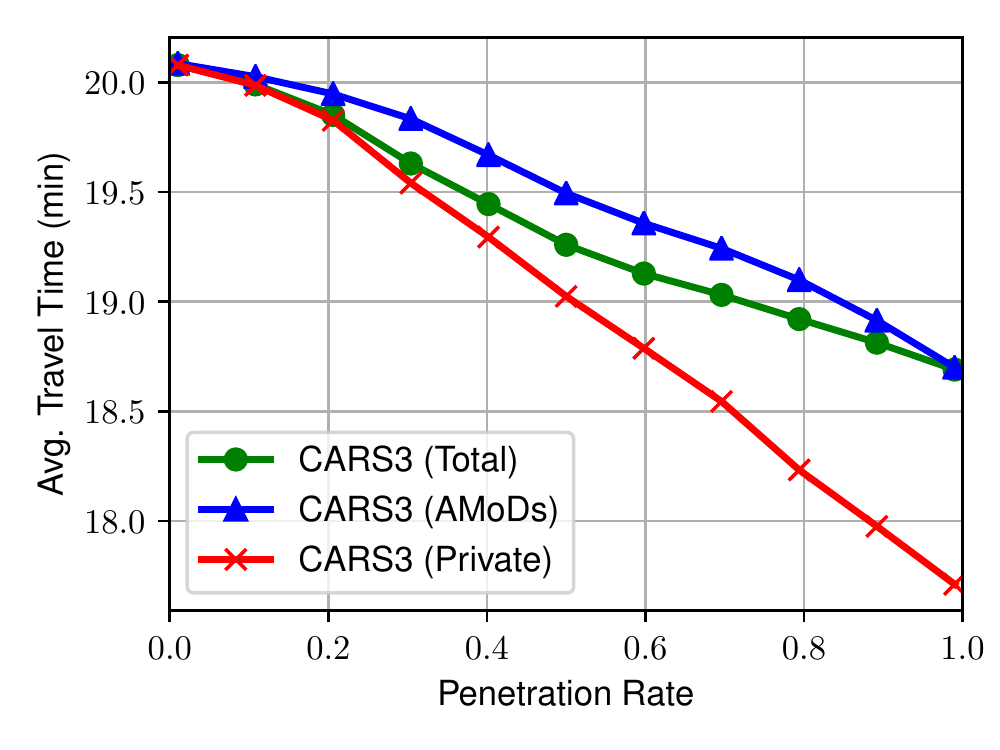}
      \caption{NYC, no rebalancing}
      \label{fig:NYC-penRate-no-rebalance}
    \end{subfigure}
    \begin{subfigure}{.465\linewidth}
      \centering
      \includegraphics[trim={0.8cm 0.4cm 0 0},clip, width=1\linewidth]{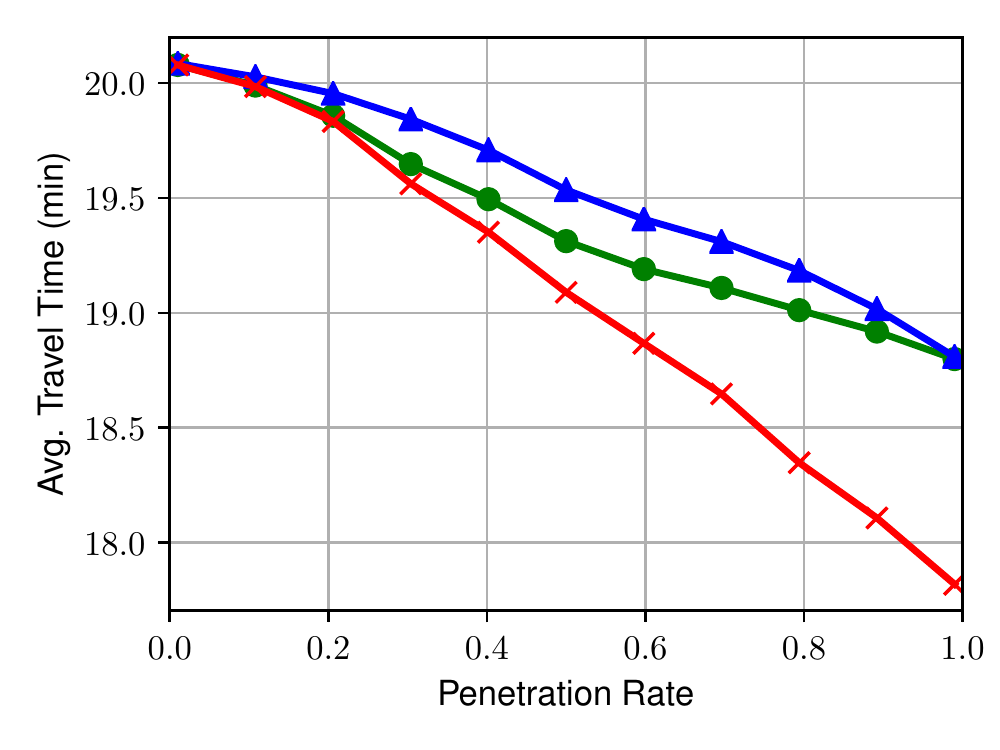}
      \caption{NYC rebalancing}
      \label{fig:NYC-penRate-rebalance}
    \end{subfigure}
    \caption{Travel times for AMoD users, private vehicles and all vehicles (total) for different penetration rates of AMoDs in the network.}
    \label{fig:rebalance-tradeoff}
    \vspace{-.45cm}
\end{figure}
\begin{figure}[t]
    \centering
    \begin{subfigure}{1\linewidth}
      \centering
      \includegraphics[trim={0cm 0.4cm 0 0},clip, width=1\linewidth]{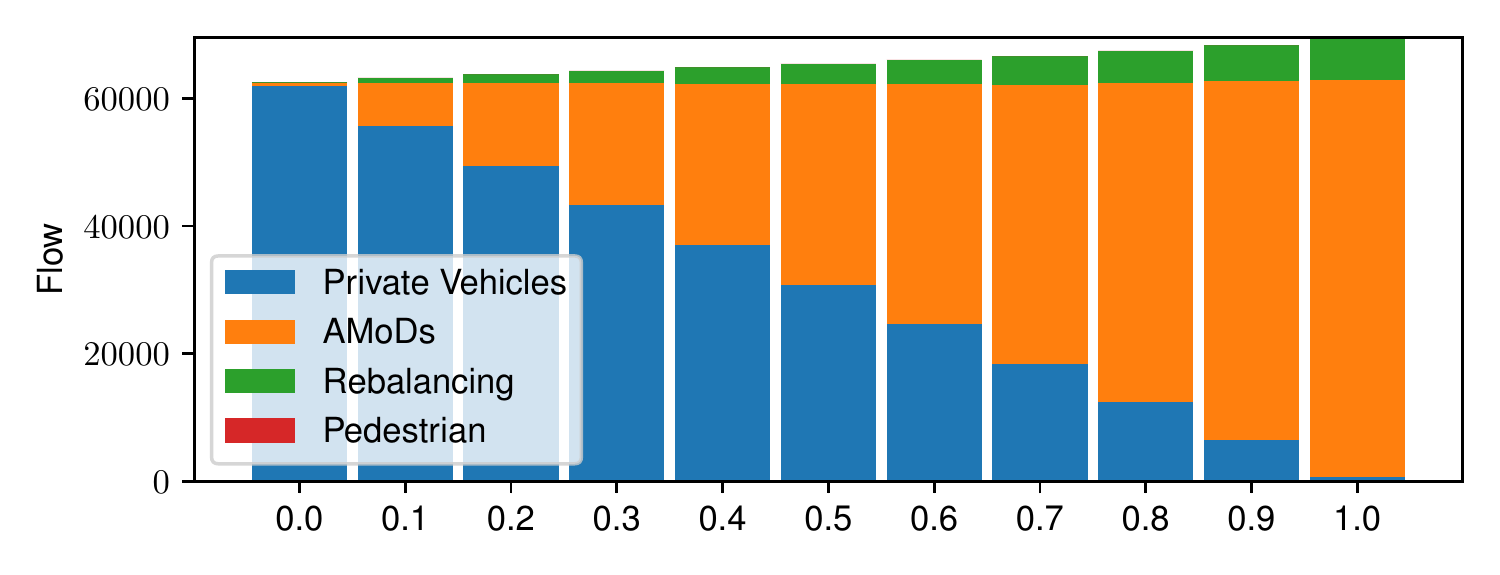}
      \label{fig:EMA-penRate-no-rebalanceds}
    \end{subfigure}
    \vspace{-0.5cm}
    \caption{Distribution of flow in EMA per mode of transport. The x-axis represents the penetration rate of AMoD users.}
    \label{fig:flow-distribution}
    \vspace{-.45cm}
\end{figure}

As shown in Figs.~\ref{fig:EMA-penRate-no-rebalance} and \ref{fig:NYC-penRate-no-rebalance}, the introduction of AMoD users into the system not only improves the overall travel time of AMoD users themselves, but reduces the travel time of private vehicles even more. This is because smart routing decisions of AMoD vehicles reduce the traffic intensity on congested roads, which consequently allows private vehicles to travel faster.
As AMoD users begin to enter the system, we see that the average travel time per vehicle decrease compared to the uncontrolled traffic scenario. Moreover, the travel time of commuting through the fastest route (private vehicles) decreases as more AMoD users are in the system.

Fig.~\ref{fig:rebalance-tradeoff} shows the interaction between the two classes of vehicles when rebalancing is used or not.
Comparing Fig.~\ref{fig:EMA-penRate-rebalance} with Fig.~\ref{fig:EMA-penRate-no-rebalance}, we see that increasing the number AMoD users (penetration rates from $0$ to $0.5$), all vehicles decrease their travel time. However, as penetration increases ($0.5$ to $1$), a larger amount of vehicles needs to be rebalanced, resulting in a rise of travel times as the overall flow in the network increases as shown in Fig.~\ref{fig:flow-distribution}.
The EMA network achieves lower benefits by using system-centric strategies, possibly because the EMA is a highway network with less degrees of freedom in terms of routing decisions than an urban setting.
In contrast, for NYC, the impact of rebalancing is negligible, and increasing the number of AMoD users allows to reduce travel time by up to 10\%. Notably, these results are in line with the low-to-medium congestion cases in the peak hour presented in~\cite[Sec.~5.2]{RossiZhangEtAl2017}.
Finally, although the results for EMA and NYC shown in Figs.~\ref{fig:EMA-penRate-rebalance} and \ref{fig:NYC-penRate-rebalance} are not identical, they follow similar trends.
In particular, for a $100\%$ AMoD penetration, rebalancing slightly increases the overall travel times for both networks.
Yet, in general, the impact of rebalancing on the system-level performance depends on the network topology, and on the symmetry of the OD demand distribution.

\subsection{Walking and Micromobility Options} \label{subsec:expermients-walking}

\begin{figure}[t] 
    \centering
        \begin{subfigure}{1\linewidth}
        \begin{subfigure}{0.34\linewidth}
          \centering
          \includegraphics[trim={0cm 0cm 0 0},clip, width=1\linewidth]{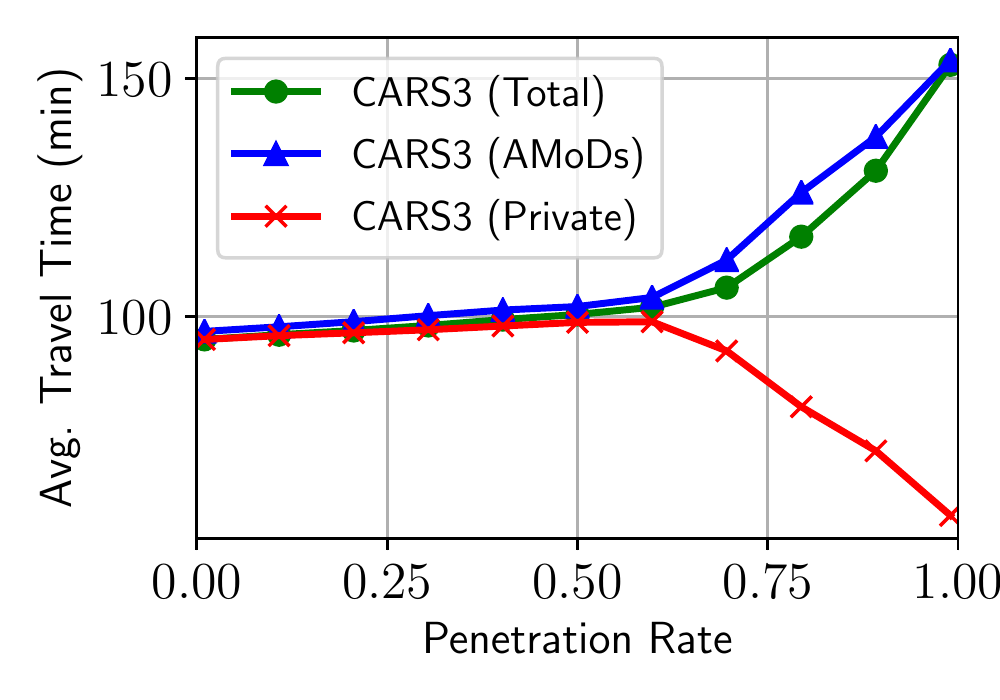}
          \vspace{-.7cm}
        \end{subfigure}
        \begin{subfigure}{0.64\linewidth}
          \centering
          \includegraphics[trim={0cm 0cm 0 0},clip, width=1\linewidth]{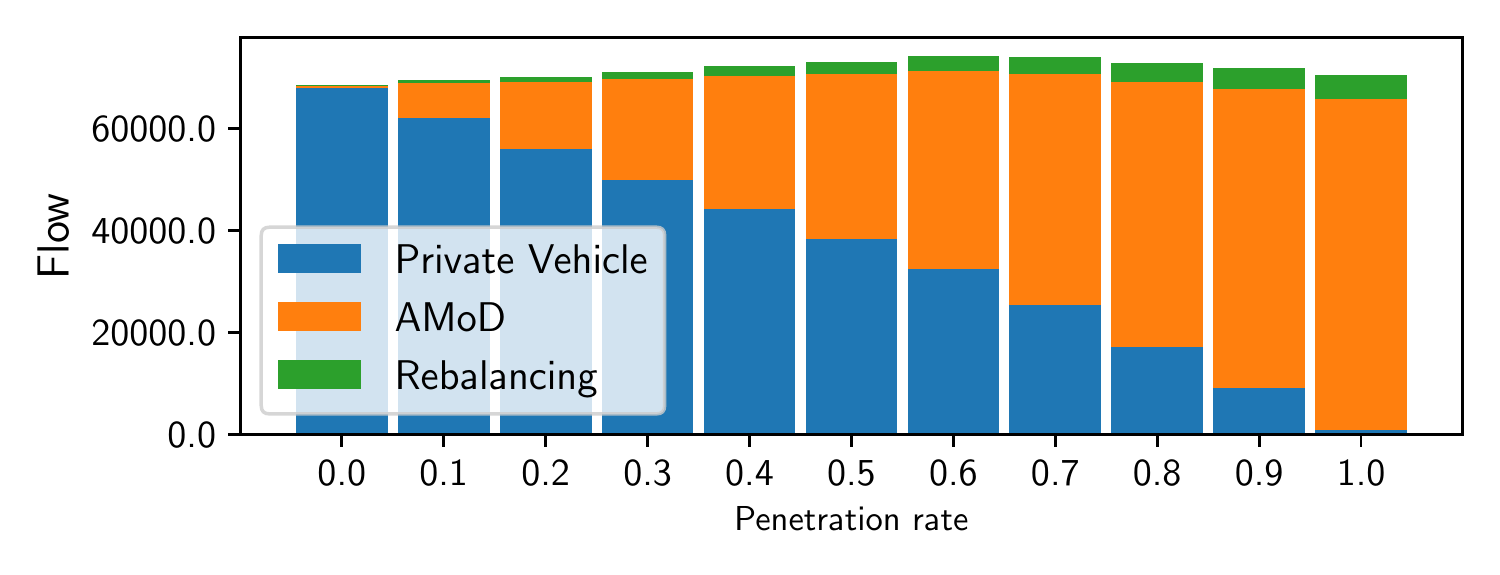}
          \vspace{-.7cm}
    \end{subfigure}
    \caption{No walking nor micromobility.}
    \label{fig:nyc-2g}
    \end{subfigure}
    \begin{subfigure}{1\linewidth}
        \begin{subfigure}{0.34\linewidth}
          \centering
          \includegraphics[trim={0cm 0cm 0 0},clip, width=1\linewidth]{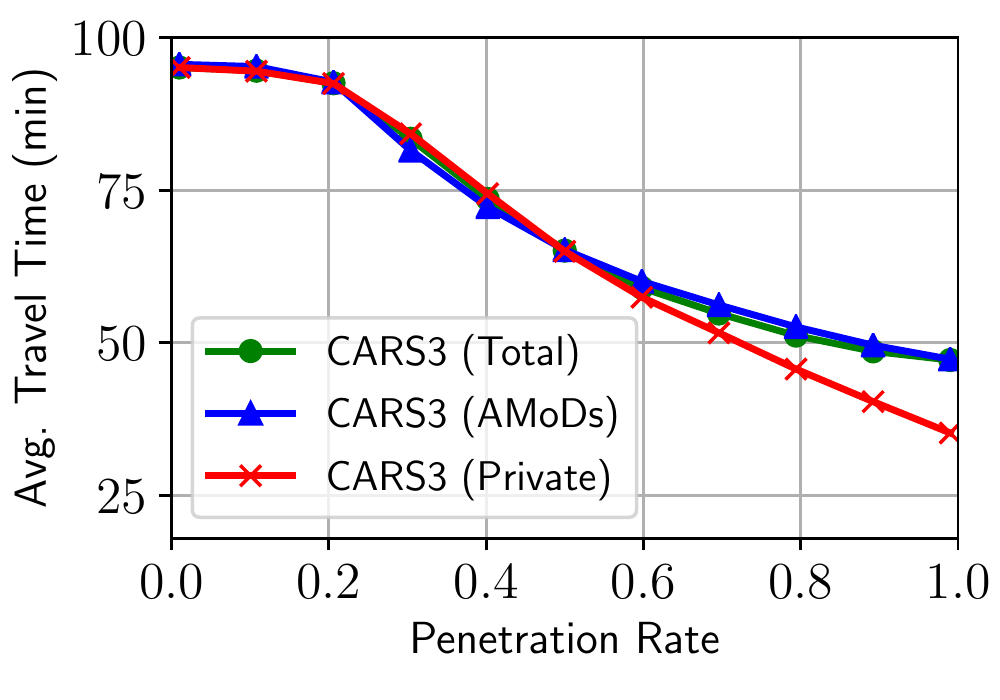}
          \vspace{-.7cm}
        \end{subfigure}
        \begin{subfigure}{0.64\linewidth}
          \centering
          \includegraphics[trim={0cm 0cm 0 0},clip, width=1\linewidth]{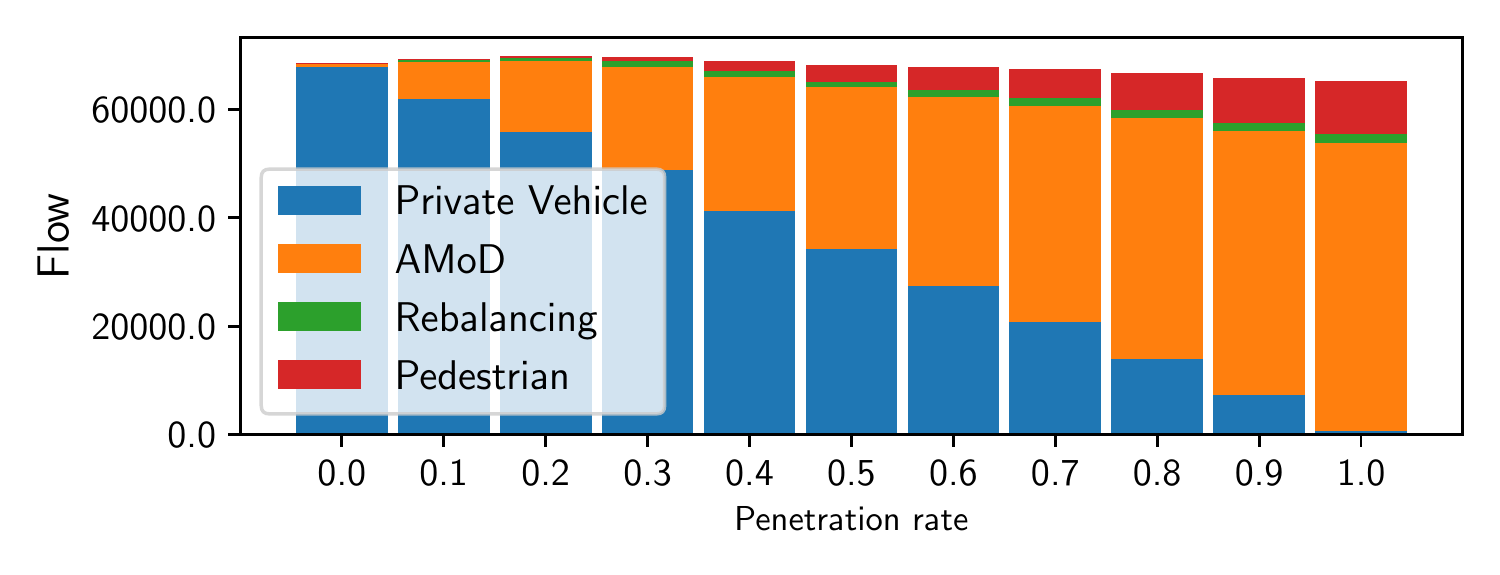}
          \vspace{-.7cm}
        \end{subfigure}
      \caption{Walking case (speed considered at \unit[3.1]{mph}).}
      \label{fig:nyc-walking}
    \end{subfigure}
    \begin{subfigure}{1\linewidth}
        \begin{subfigure}{0.34\linewidth}
          \centering
          \includegraphics[trim={0cm 0cm 0 0},clip, width=1\linewidth]{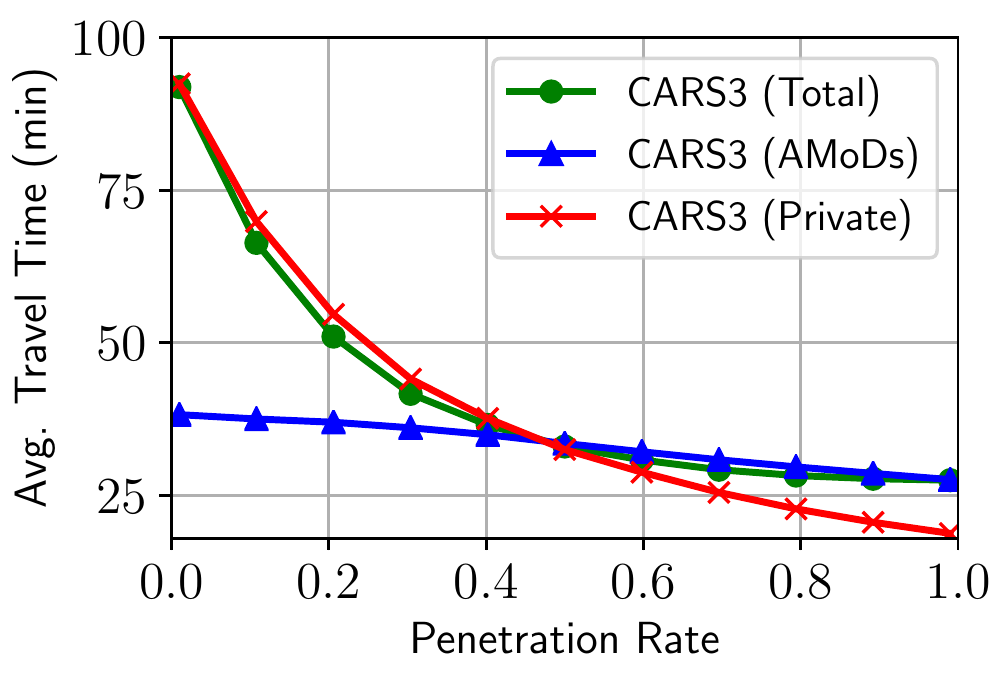}
          \vspace{-.7cm}
        \end{subfigure}
        \begin{subfigure}{0.64\linewidth}
          \centering
          \includegraphics[trim={0cm 0cm 0 0},clip, width=1\linewidth]{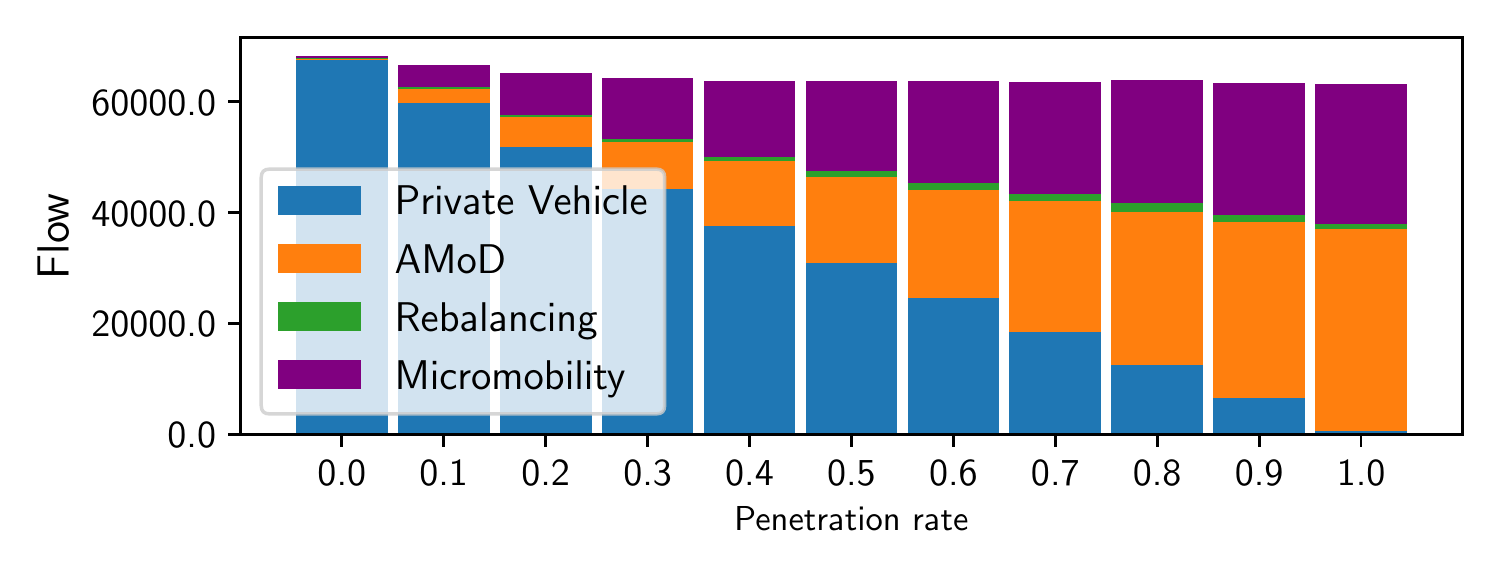}
          \vspace{-.7cm}
        \end{subfigure}
    \caption{Micromobility case (e-scooters at \unit[10]{mph}).}
    \label{fig:nyc-micromobility}
    \end{subfigure}
    \caption{Effect of alternative mode of transport in NYC when demand is high. We increase demand by a $2.5$ factor, i.e., we use $2.5\bg$.}
    \label{fig:walking-example}
    \vspace{-.6cm}
\end{figure}

In order to study the impact of centralized routing under high congestion levels, we run experiments for the NYC network with a higher overall demand level ($2.5$ times higher than in Fig.~\ref{fig:NYC-penRate-no-rebalance}).
As in the previous experiment, we run the analysis for different penetration rates.
Notably, the initial travel times shown in Fig.~\ref{fig:walking-example} are in line with the high congestion case in the peak hour in~\cite[Sec.~5.2]{RossiZhangEtAl2017}.
We observe in Fig.~\ref{fig:nyc-2g} that without considering the walking or micromobility options, the injection of AMoD users to the network increases travel times. This is a result of the additional rebalancing flow needed to operate the system in high demand periods, and happens due to the evaluation of $t(\cdot)$ at those points: Every additional flow increases travel times quartically when congestion is high.
In contrast, by leveraging the possibility of walking (Fig.~\ref{fig:nyc-walking}), the decrease in the overall travel time is much higher for higher AMoD penetration rates. In fact, for a $100\%$ penetration rate, the overall travel time is halved compared to a $0\%$ penetration rate.
Additionally, we consider the possibility of using micromobility vehicles. In particular, we analyze the case when electric scooters are available to AMoD users everywhere, for which we assume an average speed of $\unit[10]{mph}$ and the same network as the walking network $\scrG_\mathrm{W}$.
Fig.~\ref{fig:nyc-micromobility} shows that the average travel time for an AMoD user is lower than for selfish users when penetration rates are low. This happens because even for a $0\%$ penetration rate, AMoD users resort to e-scooters which are not available to private vehicles' owners. Similar to other examples, the travel times for both e-scooters and private vehicles decrease as the penetration rate increases.
In conclusion, by comparing Fig.~\ref{fig:nyc-2g} with Fig.~\ref{fig:nyc-walking} and \ref{fig:nyc-micromobility},
we see that pure AMoD systems might decrease the system-level performance due to the additional congestion resulting from rebalancing the AMoD vehicles. Yet, combining centralized-routing with the possibility of walking or using micromobility solutions such as e-scooters can significantly improve the overall travel times.

    \section{Conclusions} \label{sec:conclusions}
In this paper we studied the achievable benefits of centrally controlling an Autonomous Mobility-on-Demand (AMoD) system under mixed traffic conditions.
With the goal of minimizing the customers' travel time, we extended a previously presented quadratic model~\cite{SalazarTsaoEtAl2019} by improving its accuracy and included reactive exogenous traffic flows.
Assuming the exogenous traffic (private vehicles) to act selfishly, we leveraged an iterative method~\cite{houshmand2019penetration} to study the interaction between AMoD and private cars.
Finally, we presented numerical experiments to compare the proposed method with a disjoint strategy, and to gain insights on the achievable benefits for different AMoD penetration rates and micromobility options.
Our results showed that the proposed method outperforms the disjoint strategy in terms of computational time, and revealed that combining AMoD rides with walking and micromobility options can significantly improve the overall system-level performance.

This work can be extended as follows.
First, given the large computational time of the disjoint problem (NLP) we would like to propose a MSA-type method to solve the AMoD system-centric TAP considering exogenous flow, possibly leveraging computationally efficient algorithms such as in~\cite{SoloveySalazarEtAl2019}.
Second, we would like to generalize the approximation model to $n$ line segments, and provide theoretical bounds on the model error.
Third, given that the solution of these models are in terms of flow, we would like to include route-recovery strategies and apply this framework to larger networks through high-fidelity simulations.
Finally, we would like to consider a more general intermodal setting as in~\cite{SalazarRossiEtAl2018,SalazarLanzettiEtAl2019} by including public transportation options.

    \bibliographystyle{IEEEtran}
    \begin{tiny}
    \bibliography{references,main,ASL_papers}
    \end{tiny}
    
\end{document}